\documentclass[a4paper,fleqn,usenatbib]{mnras}

\usepackage{newtxtext,newtxmath}

\usepackage[T1]{fontenc}
\usepackage{ae,aecompl}

\usepackage{graphicx}	%
\usepackage{amsmath}	%
\usepackage{amssymb}	%
\usepackage{multicol}        %
\usepackage{bm}		%
\usepackage{pdflscape}	%
\usepackage{booktabs}	%

\newcommand{\wise}{\textit{WISE}}

\title[Resolved $\Sigma(\mathrm{H_2})$ estimator]{A new estimator of resolved molecular gas in nearby galaxies}

\author[R. Chown et al.]{
Ryan Chown,$^{1}$\thanks{E-mail: chownrj@mcmaster.ca (RC); cli2015@tsinghua.edu.cn (CL)}
Cheng Li,$^{2}$
Laura Parker,$^{1}$
Christine D. Wilson,$^{1}$
Niu Li$^{2}$
and Yang Gao$^{3}$
\\
$^{1}$Department of Physics \& Astronomy, McMaster University, Hamilton, ON L8S 4M1, Canada \\
$^{2}$Department of Astronomy, Tsinghua University, Beijing 100084, China \\
$^{3}$Purple Mountain Observatory \& Key Laboratory of Radio Astronomy, Chinese Academy of Sciences, Nanjing 210034, China
}

\date{Accepted XXX. Received YYY; in original form ZZZ}

\pubyear{2020}

\begin{document}
\label{firstpage}
\pagerange{\pageref{firstpage}--\pageref{lastpage}}
\maketitle

\defcitealias{bolatto2017}{B17}
\defcitealias{catalan-torrecilla2015}{C15}
\defcitealias{gilhuly2019}{G19}

\begin{abstract}
A relationship between dust-reprocessed light from recent star formation and the amount of star-forming gas in a galaxy produces a correlation between \wise\ 12 \micron\ emission and CO line emission. Here we explore this correlation on kiloparsec scales with CO(1-0) maps from EDGE-CALIFA matched in resolution to \wise\ 12 \micron\ images. We find strong CO-12 \micron\ correlations within each galaxy and we show that the scatter in the global CO-12 \micron\ correlation is largely driven by differences from galaxy to galaxy. The correlation is stronger than that between star formation rate and H$_2$ surface densities ($\Sigma(\mathrm{H_2})$). %
We explore multi-variable regression to predict $\Sigma(\mathrm{H_2})$ in star-forming pixels using the \wise\ 12 \micron\ data combined with global and resolved galaxy properties, and provide the fit parameters for the best estimators. We find that $\Sigma(\mathrm{H_2})$ estimators that include $\Sigma(\mathrm{12\>\mu m})$ are able to predict $\Sigma(\mathrm{H_2})$ more accurately %
than estimators that include resolved optical properties 
instead of $\Sigma(\mathrm{12\>\mu m})$. 
These results suggest that 12 \micron\ emission and H$_2$ as traced by CO emission 
are physically connected
at kiloparsec scales. This may be due to a connection between polycyclic aromatic hydrocarbon (PAH) emission and the presence of H$_2$.
The best single-property estimator is $\log \frac{\Sigma(\mathrm{H_2})}{\mathrm{M_\odot\>pc^{-2}}} = (0.48 \pm 0.01) +  (0.71 \pm 0.01)\log \frac{\Sigma(\mathrm{12\>\mu m})}{\mathrm{L_\odot\>pc^{-2}}}$. 
This correlation can be used to efficiently estimate $\Sigma(\mathrm{H_2})$ down to at least $1 \> M_\odot \> \mathrm{pc^{-2}}$ in star-forming regions within nearby galaxies.
\end{abstract}


\section{Introduction}

Stars form out of molecular hydrogen  
in cold, dense regions of the interstellar medium (ISM).
Empirically this picture is supported by correlations between
tracers of cold gas and the radiation output from young stars
such as the Kennicutt-Schmidt (KS) law 
\begin{equation} \label{eq:ks}
\Sigma(\mathrm{SFR}) \propto \Sigma(\mathrm{gas})^N,
\end{equation}
where $\Sigma(\mathrm{SFR})$ is the star formation rate (SFR) 
surface density ($M_\odot \> \mathrm{kpc}^{-2}$),
$\Sigma(\mathrm{gas})$ is the atomic (\ion{H}{I}) + molecular (H$_2$) gas
surface density ($M_\odot \> \mathrm{pc}^{-2}$), and $N$ is a power-law
index of $\simeq1.4$, or $\simeq1.0$ if only H$_2$ is included 
\citep{kennicutt1989, kennicutt2007, bigiel2008, leroy2008, leroy2013}.
Within the scatter of the KS 
law, there are systematic variations between galaxies and sub-regions within
galaxies, suggesting that this law may not be universal
\citep{shetty2013}. 
For instance, below $\Sigma(\mathrm{gas})\simeq10 \> M_\odot \> \mathrm{pc^{-2}}$ 
and $\Sigma(\mathrm{SFR}) \lesssim 10^{-3} \> M_\odot\mathrm{yr^{-1}kpc^{-2}}$, 
the stellar mass surface density $\Sigma_*$ becomes important
in regulating 
the star formation rate ($\Sigma(\mathrm{SFR})\propto [\Sigma_*^{0.5}\Sigma(\mathrm{gas})]^{1.09}$) \citep{shi2011, shi2018}.
Another example of a modification to the KS law is the Silk-Elmegreen law,
which incorporates the orbital dynamical timescale $\Sigma(\mathrm{SFR})\propto t_\mathrm{dyn}^{-1}\Sigma(\mathrm{gas})$ \citep{elmegreen1997, silk1997}. 
On the galaxy-integrated (``global'') side, 
\citet{gao2004} found a strong correlation between 
global measurements of HCN luminosity (a dense molecular gas tracer) 
and total infrared luminosity (a SFR tracer) 
ranging from normal spirals to ultraluminous infrared galaxies, again
supporting a picture in which stars form in cold dense gas.
The physical interpretation 
of these relationships requires an understanding of the 
limitations and mechanisms behind the tracers used to 
measure $\Sigma(\mathrm{SFR})$ and $\Sigma(\mathrm{gas})$ \citep[e.g.][]{krumholz2007}.

One manifestation of the KS law is the correlation between 
12 \micron\ luminosity, measured with the \textit{Wide-field Infrared Survey Explorer} 
\citep[\wise; ][]{wright2010}, and CO luminosity measured by 
ground-based radio telescopes.
The 12 \micron\ (also called W3) band spans mid-infrared (MIR) wavelengths of 8 to 16 \micron. 
In nearby galaxies, 12 \micron\ emission 
traces SFR 
\citep[e.g.][]{donoso2012, jarrett2013, salim2016, cluver2017, salim2018, leroy2019},
vibrational emission lines from polycyclic aromatic hydrocarbons (PAHs), and 
warm dust emission \citep{wright2010}.
PAHs are excited primarily by stellar UV emission via the photoelectric effect, and the main features
appear at wavelengths of 3.3, 6.2, 7.7, 8.6, 11.3, 12.7 and 16.4 \micron~\citep{bakes1994, tielens2008}.
Where and how PAHs form is a topic of ongoing debate, but PAH emission is associated with star formation \citep[e.g.][]{peeters2004, xie2019, whitcomb2020}
as well as CO emission \citep[e.g.][]{regan2006, sandstrom2010, pope2013, cortzen2019, li2020a}.
Galaxy-integrated 12 \micron\ luminosity 
is strongly correlated with CO(1-0) and CO(2-1) luminosity 
in nearby galaxies \citep{jiang2015, gao2019}.
\citet{gao2019} find
\begin{equation}
\log \left(\frac{L_\mathrm{CO(1-0)}}{\mathrm{K\>km\>s^{-1}\>pc^2}}\right) = N \log \left(\frac{L_\mathrm{12\>\mu m}}{\mathrm{L_\odot}}\right) + \log C,
\end{equation}
with $N = 0.98 \pm 0.02 $ and $\log C = -0.14 \pm 0.18$, 
and scatter of 0.20 dex.
The correlation between \wise\ 22 \micron\ 
luminosity, which is dominated by warm dust emission, and CO luminosity is weaker (0.3 dex scatter)
than that between 12 \micron\ and CO (0.2 dex scatter), implying 
that 12 \micron\ luminosity is a better indicator of CO luminosity than 
22 \micron\ \citep{gao2019}.
Since the prominent 11.3 \micron\ PAH feature lies in the \wise\ 12 \micron\ band,
it is possible that the 12 \micron-CO correlation is strengthened by a combination of 
the Kennicutt-Schmidt relation (since PAH emission traces SFR) and the link
between CO emission and PAH emission.
The scatter in the global 12 \micron-CO fit is reduced to 0.16 dex when 
$g-r$ colour and stellar mass
are included as extra variables in the fit \citep{gao2019}.
Empirical relationships such as these are useful for predicting 
molecular gas masses in galaxies, since 12 \micron\ images are easier to 
obtain than CO luminosities. 
Mid-infrared tracers of cold gas will 
be particularly useful upon the launch of the 
\textit{James Webb Space Telescope}, which will 
observe the MIR sky with better resolution and sensitivity
than \wise.

Optical extinction $A_V$ estimated from the Balmer decrement $\mathrm{H\alpha/H\beta}$ 
has also been used
as an H$_2$ mass tracer in nearby galaxies \citep{guver2009, barrera-ballesteros2016, concas2019, yesuf2019, barrera-ballesteros2020, yesuf2020}.
The correlation between extinction (measured either by 
stellar light absorption $A_V$ or gas absorption $\mathrm{H\alpha/H\beta}$) and H$_2$ is due to the correlation between dust and H$_2$.
This method is convenient since
spatially resolved extinction maps are available for large samples of galaxies
thanks to optical integral-field spectroscopy surveys.
However, unlike 12 \micron,
extinction as measured by the Balmer decrement
is only valid over a range that is limited by the signal-to-noise ratio 
of the H$\beta$ emission line. With extreme levels of extinction, e.g. in 
local ultra-luminous infrared galaxies, the H$\beta$ line 
becomes invisible, so this method cannot be used.

It is not yet known whether the correlation between 
12 \micron\ and CO holds at sub-galaxy scales, or how 
it compares with the resolved SFR-H$_2$ and $A_V$-H$_2$ correlations.
Comparing these correlations at resolved scales may give insight into 
the factors driving the 12 \micron-CO correlation.
The \wise\ 12 \micron\ beam full-width at half-maximum (FWHM) is 6.6 arcsec \citep{wright2010}, which corresponds to $\leq 1$ kpc 
resolution for galaxies closer than 31 Mpc.
This resolution and distance range is well-matched to 
the Extragalactic Database for Galaxy Evolution survey \citep[EDGE; ][]{bolatto2017}. 
EDGE is a survey of CO(1-0) in 126 nearby galaxies with $4.5$ arcsec spatial resolution
using the Combined Array for Research in Millimeter-wave Astronomy (CARMA). One 
of the main goals of EDGE was to allow 
studies of resolved molecular gas and optical integral-field spectroscopy data in a
large sample of nearby galaxies.

In this study, we use the EDGE CO and \wise\ data to measure the 12 \micron\ and CO(1-0) 
correlation
within individual galaxies.
We find that the best-fit parameters describing this relation
vary significantly among galaxies.
We perform multivariate linear regression using
a combination of global galaxy measurements and 
quantities derived from spatially 
resolved optical spectroscopy from the 
Calar Alto Legacy Integral Field Area Survey \citep[CALIFA; ][]{sanchez2012, walcher2014, sanchez2016}.
This yields a set of linear functions with $\log\Sigma(\mathrm{H_2})$ as the 
dependent variable which can be used as spatially resolved estimators 
of H$_2$ surface density.
These estimators can predict H$_2$ surface density with an RMS accuracy of $\simeq 0.2$ dex in
galaxies for which 12 \micron\ data are available.

\begin{figure*}
	\includegraphics[width=2\columnwidth]{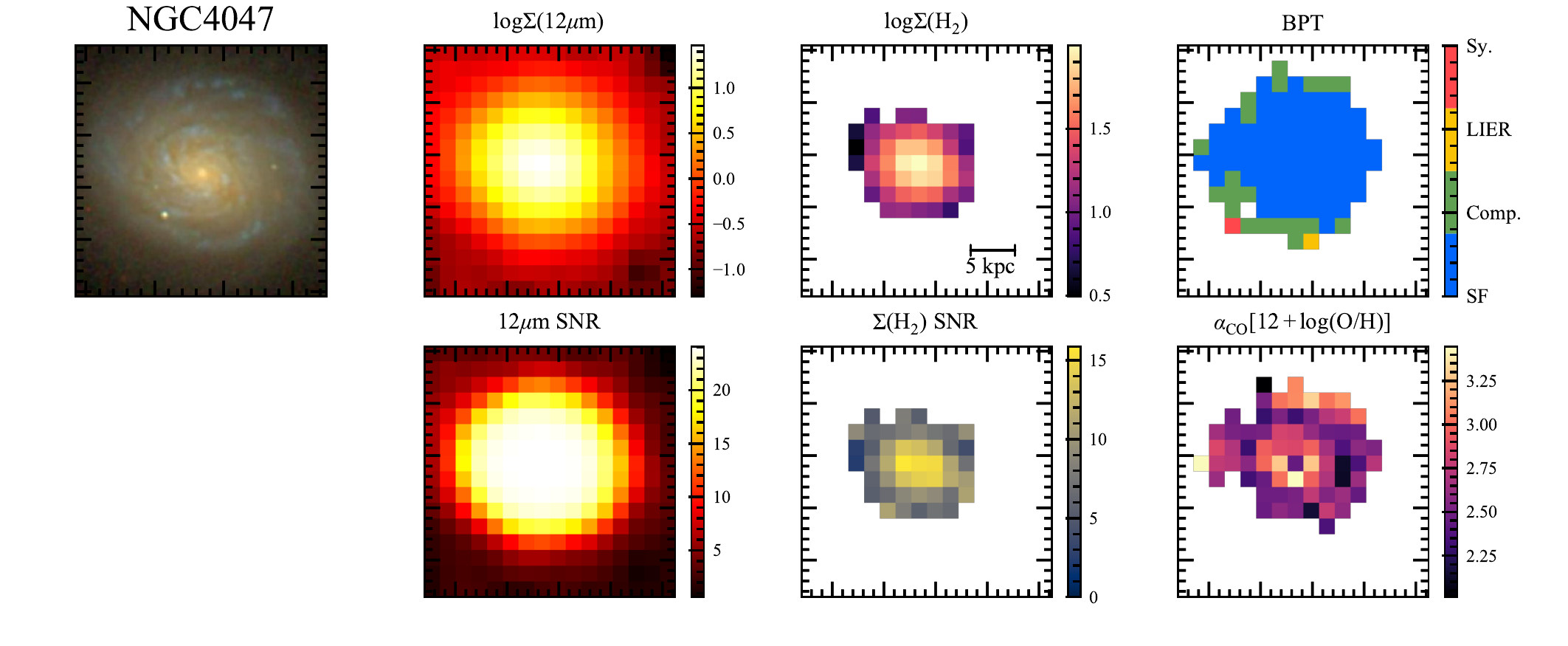}
    \caption{Selected maps for an example galaxy. Top row (left to right): Sloan Digital Sky Survey \citep[SDSS; ][]{blanton2017} \textit{gri} thumbnail; \wise\ 12 \micron\ surface
    density ($L_\odot \> \mathrm{pc^{-2}}$); H$_2$ mass surface density ($M_\odot \> \mathrm{pc^{-2}}$) at 6.6 arcsec resolution and assuming $\alpha_\mathrm{CO}=3.2$ $M_\odot \> \mathrm{(K\>km\>s^{-1}\>pc^2)}^{-1}$; BPT diagram for each pixel
    constructed from the processed CALIFA data (Section~\ref{sec:califa}). The pixel size is 6 arcsec, and the cutouts are 96-by-96 arcsec.
    Bottom row: signal-to-noise ratio (SNR) of the 12 \micron\ and H$_2$ surface density maps, and the metallicity-dependent
    $\alpha_\mathrm{CO}$ values in units of $M_\odot \> \mathrm{(K\>km\>s^{-1}\>pc^2)}^{-1}$ (Equation~\ref{eq:alphaco_met}).}
    \label{fig:example_gal}
\end{figure*}

\section{Data and Data Processing}

\subsection{Sample selection}

The sample is selected from the EDGE survey \citep[][hereafter~\citetalias{bolatto2017}]{bolatto2017}.
The typical angular resolution of EDGE CO maps is 
4.5 arcsec, and the typical H$_2$ surface density 
sensitivity before deprojecting galaxy inclination is 11 $M_\odot$ pc$^{-2}$~\citepalias{bolatto2017}.
Every EDGE galaxy has optical integral field unit (IFU) data from CALIFA, allowing 
joint studies of the content and kinematics of
cold gas (H$_2$), ionized gas,  
and stellar populations, all with $\sim$kpc spatial resolution.
We processed the CO data for all 126 EDGE galaxies, and as a starting point we selected 
the 95 galaxies which had at least one detected pixel after smoothing to 6.6 arcsec resolution and
regridding the moment-0 maps with 6 arcsec pixels (Section~\ref{sec:h2maps}).
We then selected those galaxies with inclinations less than 75 degrees, leaving
83 galaxies.
Inclination angles were derived from CO rotation curves where available~\citepalias{bolatto2017}, 
and otherwise were taken from the HyperLEDA database \citep{makarov2014}. 
Redshifts $z$ (from CALIFA emission lines) 
and luminosity distances $D_L$ 
were taken from~\citetalias{bolatto2017}.
A flat $\Lambda$CDM cosmology was assumed ($h=0.7$, $\Omega_m=0.27$, $\Omega_\Lambda=0.73$).

\subsection{\wise\ 12\micron\ surface density maps}

We downloaded 2 degree by 2 degree
cutouts (pixel size 1.375 arcsec) 
of \wise\ 12 \micron\ (W3) flux $F_\mathrm{W3}$ and uncertainty 
for each galaxy
from the NASA/IPAC Infrared Science Archive.
The background for each galaxy was estimated using the IDL package Software for Source Extraction \citep[SExtractor;][]{bertin1996},
with default parameters and with the corresponding W3 uncertainty map as input.
The estimated background was subtracted from each cutout. The background-subtracted
images were reprojected with 6 arcsec pixels to avoid over-sampling the 6.6 arcsec beam.
These maps were originally in units of Digital Numbers (DN), defined such that a W3 magnitude $m_\mathrm{W3}$ of 18.0 
corresponds to $F_\mathrm{W3} = 1.0$ DN, or
\begin{equation}
F_\mathrm{W3} = 10^{-0.4(m_\mathrm{W3}-\mathrm{MAGZP})} \> \mathrm{DN},
\end{equation}
where the zero-point magnitude $\mathrm{MAGZP}=18.0$ mag.
We converted the maps from their original units to 
flux density in Jy, given by
\begin{align}
S_\mathrm{W3} &= S_0 10^{-0.4m_\mathrm{W3}}\\
&= S_0 10^{-0.4\mathrm{MAGZP}} F_\mathrm{W3}\\
&= \left(\frac{31.674}{10^{7.2}} \> \mathrm{Jy\> DN^{-1}}\right)F_\mathrm{W3}\\
&= \left(1.998\times 10^{-6} \> \mathrm{Jy\> DN^{-1}}\right)F_\mathrm{W3},
\end{align}
where the isophotal flux density $S_0 = 31.674$ Jy for the W3 band 
is from Table 1 of \citet{jarrett2011}. 
Luminosity in units of $\mathrm{L_\odot}$ is given by 
\begin{align}
L_\mathrm{12\>\mu m} &= 4\pi D_L^2 \Delta \nu S_\mathrm{W3}\\
&= 7.042 F_\mathrm{W3} \left(\frac{D_L}{\mathrm{Mpc}}\right)^2 \> \mathrm{L_\odot}
\end{align}
where $\Delta \nu = 1.1327 \times 10^{13}$ Hz is the bandwidth of 
the 12 \micron\ band \citep{jarrett2011}, and
$D_L$ is the luminosity distance.
Luminosities were then converted into surface densities $\Sigma(\mathrm{12\>\mu m})$ 
($\mathrm{L_\odot}$ pc$^{-2}$) by
\begin{equation}\label{eq:w3units}
\frac{\Sigma(\mathrm{12\>\micron})}{\mathrm{L_\odot} \> \mathrm{pc}^{-2}} = 7.042 \left(\frac{F_\mathrm{W3}}{\mathrm{DN}}\right) \left(\frac{D_L}{\mathrm{Mpc}}\right)^2 \left(\frac{A_\mathrm{pix}}{\mathrm{pc^2}}\right)^{-1}\cos i,
\end{equation}
where $i$ is the galaxy inclination, and
$A_\mathrm{pix}$ is the pixel area in pc$^2$.

The uncertainty in each pixel of the rebinned surface density maps is the quadrature sum of 
the instrumental uncertainty and the 4.5 per cent uncertainty in the zero-point magnitude (Appendix~\ref{appendix:w3unc}). Maps for an example galaxy are shown in 
Figure~\ref{fig:example_gal}.

\subsection{H$_2$ surface density maps at \wise\ W3 resolution}\label{sec:h2maps}

The original CO(1-0) datacubes were downloaded from the EDGE website,\footnote{\url{https://mmwave.astro.illinois.edu/carma/edge/bulk/180726/}}
converted from their native units of K km s$^{-1}$ to $\mathrm{Jy\>beam^{-1}\>km\>s^{-1}}$,
and then smoothed to a Gaussian beam with FWHM $=6.6$ arcsec using the
Common Astronomy Software Applications \citep[CASA; ][]{mcmullin2007} task \verb'imsmooth' to match the \wise\ resolution.
The cubes have a velocity resolution of 20 km s$^{-1}$, and span 44 channels (880 km s$^{-1}$).
Two methods were used to obtain 
CO integrated intensity (moment-0) maps $S_\mathrm{CO}\Delta v$: 
\begin{enumerate}
	\item[\textbf{Method 1:}] an iterative masking technique for improving SNR, described in \citet{sun2018}, shown in Figure~\ref{fig:example_gal}, and
	\item[\textbf{Method 2:}] integrating the flux along the inner 34 channels (680 km s$^{-1}$ total). In this ``simple'' method, the first 5 and last 5 channels were used to compute the root-mean-square (RMS) noise at each pixel.
\end{enumerate}
Method 1 is used for all results in this work, 
while Method 2 is used as a cross-check and to estimate upper limits for non-detected pixels.

In Method 1 \citep[described in][]{sun2018} a mask is generated for the 
datacube to improve the signal-to-noise
of the resulting moment-0 map.
A ``core mask'' is generated by requiring 
SNR of 3.5 over 2 consecutive channels (channel width of 20 km s$^{-1}$),
and a ``wing mask'' is generated by requiring SNR of 2.0 over 2 consecutive 
channels. The core mask is dilated within the wing mask to generate a ``signal mask''
which defines detections. Any detected regions that span an area less than 
the area of the beam are masked. The signal mask is then extended spectrally 
by $\pm 1$ channels. Method 2 gives a map with lower signal-to-noise, but is useful for 
computing upper-limits for pixels which are masked in Method 1, and 
for cross-checking results.

The moment-0 maps were then rebinned with 6 arcsec pixels, and the units were
converted to integrated intensity per pixel %
\begin{equation}
\frac{S_\mathrm{CO}\Delta v}{\mathrm{Jy\>km\>s^{-1}\>pixel^{-1}}} = \left(\frac{S_\mathrm{CO}\Delta v}{\mathrm{Jy\> beam^{-1} \> km \> s^{-1}}}\right) \frac{4\theta_\mathrm{pix}^2\ln 2}{\pi \mathrm{FWHM}^2},
\end{equation}
where the beam $\mathrm{FWHM}=6.6$ arcsec, and the pixel
size $\theta_\mathrm{pix}=6$ arcsec.

The total noise variance 
in each pixel is the sum in quadrature of the instrumental noise 
which we assume to be the same for both moment-0 map versions, 
and calibration uncertainty which depends on the moment-0 method (Appendix~\ref{appendix:counc}).
Instrumental noise maps 
were computed by measuring the 
RMS in the first 
five and final five channels at each pixel (Method 2 above). The instrumental noise maps
were rebinned (added in quadrature, then square root) into 6 arcsec pixels.
To obtain the \textit{total} noise for each moment-0 map,
a calibration uncertainty of 10 per cent~\citepalias{bolatto2017} of the rebinned moment-0 map (both versions described above)
was added in quadrature with the 
instrumental uncertainty.
The sensitivity of the CO data is worse than that of \wise\ W3, and so
upper limits for undetected pixels are calculated with the second moment-0 map-making method.
All pixels detected at less than 3$\sigma$ in CO were assigned 
an upper limit of 5 times the noise at each pixel.

The CO(1-0) luminosity and noise maps (in units of K km s$^{-1}$ pc$^2$) 
were computed via \citep{bolatto2013}
\begin{equation}\label{eq:lco1}
L_\mathrm{CO(1-0)} = \frac{2453(S_\mathrm{CO}\Delta v)D_L^2}{1+z},
\end{equation}
where $z$ is the redshift.
The luminosity maps were converted to H$_2$-mass surface density $\Sigma (\mathrm{H_2})$
using a CO-to-H$_2$ conversion factor $\alpha_\mathrm{CO}$
\begin{equation}\label{eq:sigh2}
\Sigma (\mathrm{H_2}) = \frac{\alpha_\mathrm{CO}L_\mathrm{CO}\cos i}{A_\mathrm{pix}},
\end{equation}
where $i$ is the galaxy inclination angle, and $A_\mathrm{pix}$ is the pixel area in pc$^2$.
In normal star-forming regions a CO-to-H$_2$ conversion factor
of $\alpha_\mathrm{CO} = 3.2 \> \mathrm{M_\odot (K\> km \> s^{-1} \> pc^2)^{-1}}$
(multiply by 1.36 to include helium) is often assumed \citep{bolatto2013}.
We consider both a constant $\alpha_\mathrm{CO}$ and a spatially-varying 
metallicity-dependent 
$\alpha_\mathrm{CO}$ (Section~\ref{sec:aco}).

\subsection{Maps of stellar population and ionized gas properties} \label{sec:califa}

In the third data release (DR3) of the CALIFA survey there are 667 galaxies 
observed out to at least two effective radii with $\simeq2.5$ arcsec 
angular resolution over wavelengths 3700-7500~\AA\ \citep{sanchez2012, sanchez2016}. 
The observations were carried out in either a medium spectral resolution mode 
(``$V_{1200}$,'' $R\simeq1700$, 3700-4200 \AA, 484 galaxies) or a low spectral resolution mode 
(``$V_{500}$,'' $R\simeq850$, 3750-7500 \AA, 646 galaxies).
Cubes using data from both $V_{1200}$ and $V_{500}$ were made by degrading the spectral
resolution of the $V_{1200}$ cube to that of $V_{500}$ and averaging the spectra where their wavelength coverage overlaps, and using only 
$V_{1200}$ or $V_{500}$ for the remaining wavelength bins between 
3700-7140 \AA\ \citep{sanchez2016}.
Combined $V_{1200}+V_{500}$ datacubes and $V_{500}$ datacubes were downloaded from 
the CALIFA DR3 webpage.\footnote{\url{https://califaserv.caha.es/CALIFA_WEB/public_html/?q=content/califa-3rd-data-release}}
Of the 95 EDGE galaxies detected in CO,
combined $V_{1200}+V_{500}$ datacubes are available for 87 galaxies. 
$V_{500}$ datacubes were used for the remaining 8 galaxies.
We refer to this sample of 8 + 87 galaxies as ``Sample A''  (Table~\ref{tab:selection}).

\begin{table*}
	\centering
	\caption{Summary of the number of pixels and galaxies at each stage of sample selection.
	Note that Samples B and C are selected from Sample A. Sample C is the starting point for Section~\ref{sec:bayesfit} onwards.}
	\label{tab:selection}
	\begin{tabular}{clrrr} %
		\hline
		Sample label & Criteria & \#\ pixels & \#\ galaxies & Where used\\
		\hline
		A & At least one CO-detected pixel,$^\star$ and have $V_{500}+V_{1200}$ or just $V_{500}$ CALIFA datacubes & 2317$^\dagger$ & 95 & \\
		B & A $\cap$ Have at least 4 CO-detected pixels per galaxy and inclination $i<75\deg^{\ddagger}$ & 2059 & 83 & Figures~\ref{fig:pearson_hist},~\ref{fig:w3_vs_co_v2} \\
		C &  A $\cap$ Have at least 4 CO-detected pixels classified as star-forming per galaxy and $i<75\deg$ & 1168 & 64 & Figures~\ref{fig:sigma12_sigmah2_overall},~\ref{fig:pearson_hist},~\ref{fig:slope_int_summary},~\ref{fig:w3_vs_co},~\ref{fig:w3_vs_co_n} \\
		\hline
\multicolumn{5}{l}{$^\star$ Using Method 1 (Section~\ref{sec:h2maps}).}\\
\multicolumn{5}{l}{$^\dagger$ CO-detected pixels only.}\\
\multicolumn{5}{l}{$^{\ddagger}$ The reduction in the number of pixels and galaxies when going from Sample A to Sample B is entirely from the inclination cut.}\\
	\end{tabular}
\end{table*}

The native pixel size of a CALIFA cube is 1 arcsec. 
The spaxels were stacked into 6 arcsec spaxels to be compared with
the \wise\ and EDGE CO data.
Spectral fitting was performed on the stacked spectra using the Penalized Pixel-Fitting (pPXF) Python package 
\citep{cappellari2017} to obtain 2D maps of emission and absorption line fluxes, 
equivalent widths, and velocity dispersions, as well as stellar population 
properties such as stellar mass and light-weighted stellar age.
A Kroupa initial mass function (IMF) was assumed \citep{kroupa2003}.

Line fluxes were corrected for extinction using the Balmer decrement.
Stellar mass was measured from the datacubes after 
subtracting a dust extinction curve using the method of \citet{li2020}. 
The unattenuated H$\alpha$ emission line 
flux $F_\mathrm{H\alpha}$ is related to the observed (attenuated) flux according to
\begin{equation}
F_\mathrm{H\alpha} = F_\mathrm{H\alpha,obs.} 10^{0.4 A_V}
\end{equation}
where the extinction is given by
\begin{equation}\label{eq:a_ha}
A_V = 5.86 \log \left(\frac{F_\mathrm{H\alpha,obs.}}{2.86 F_\mathrm{H\beta,obs.}}\right),
\end{equation}
and $F_\mathrm{H\alpha,obs.}$ and $F_\mathrm{H\beta,obs.}$ are the observed (attenuated) 
line fluxes.
The star formation rate (SFR) surface density is given by 
\begin{align}\label{eq:sigma_sfr}
\Sigma(\mathrm{SFR}) &= \frac{C_\mathrm{SFR,H\alpha}L_\mathrm{H\alpha} }{A_\mathrm{pix}}\\
&=\frac{C_\mathrm{SFR,H\alpha} F_\mathrm{H\alpha} 4\pi d^2 \cos i}{A_\mathrm{pix}},
\end{align}
where the H$\alpha$ luminosity-to-SFR calibration factor $C_\mathrm{SFR,H\alpha}=5.3\times 10^{-42} \frac{M_\odot\>\mathrm{yr}^{-1}}{\mathrm{erg\>s^{-1}}}$ \citep{hao2011, murphy2011, kennicutt2012},
$d$ is the luminosity distance in cm, and $A_\mathrm{pix}$ is the pixel area in kpc$^2$.

The mechanism of gas ionization  
at each pixel was classified as either 
star formation (SF), low-ionization emission region (LIER),
Seyfert (Sy) or a combination of star formation and AGN (``composite'')
on a Baldwin, Phillips, and Terlevich (BPT) diagram \citep{baldwin1981}.
It is important to identify non-starforming
regions, especially when estimating SFR from H$\alpha$ flux.
BPT classification (Figure~\ref{fig:example_gal}) was done in the 
[\ion{O}{III}] $\lambda 5007$/H$\beta$ vs. [\ion{N}{II}] $\lambda 6584$/H$\alpha$ plane 
using three standard demarcation curves in this space:
Eq. 5 of \citet{kewley2001}, Eq. 1 of \citet{kauffmann2003}, 
and Eq. 3 of \citet{cid-fernandes2010} \citep[see Figure 7 of][]{husemann2013}.

\subsection{CO-to-H$_2$ conversion factor}\label{sec:aco}

The CO-to-H$_2$ conversion factor $\alpha_\mathrm{CO}$ increases slightly with 
decreasing metallicity \citep{maloney1988, wilson1995, genzel2012, bolatto2013}. 
At lower metallicities, and consequently 
lower dust abundance \citep{draine2007} and
dust shielding, CO is preferentially photodissociated relative to H$_2$.
This process leads to an increase in $\alpha_\mathrm{CO}$ \citep{bolatto2013}.

A metallicity-dependent $\alpha_\mathrm{CO}$ equation \citep{genzel2012} was 
calculated at each star-forming pixel (Figure~\ref{fig:example_gal})
\begin{equation}\label{eq:alphaco_met}
\log \left(\frac{\alpha_\mathrm{CO}}{\mathrm{M_\odot} \mathrm{(K\> km \> s^{-1} \> pc^2)^{-1}}}\right) = a + b[12+\log (\mathrm{O/H})],
\end{equation}
where $a=12\pm 2$, and $b=-1.30\pm 0.25$. 
Gas-phase metallicity $12+\log (\mathrm{O/H})$ was computed for the star-forming pixels using
\begin{equation} \label{eq:metallicity}
12+\log (\mathrm{O/H}) =p + q\log \left(\frac{\mathrm{[\ion{N}{II}]} \> \lambda 6584}{\mathrm{H\alpha}}\right),
\end{equation}
where $p=9.12 \pm 0.05$, and $q=0.73 \pm 0.10$  \citep{denicolo2002}.
Following other works that have used this $\alpha_\mathrm{CO}(Z)$ 
relation \citep[e.g.][]{genzel2015, tacconi2018, bertemes2018}, we consciously choose
not to include the uncertainty on $\alpha_\mathrm{CO}(Z)$ (which comes from the uncertainties in $a$, $b$, $p$, and $q$) in our analysis, so that   
the uncertainties on $\log \Sigma(\mathrm{H_2})$ only reflect measurement and calibration uncertainties 
and not systematic uncertainties in the conversion factor.

The metallicity-dependent $\alpha_\mathrm{CO}=\alpha_\mathrm{CO}(Z)$ (Eq.~\ref{eq:alphaco_met})
is our preferred $\alpha_\mathrm{CO}$ because it is the most physically accurate.
This choice of $\alpha_\mathrm{CO}$ has two effects on the sample: 
\begin{enumerate}
\item the exclusion of non-starforming pixels; and
\item galaxies that have fewer
star-forming pixels with CO detections than a given threshold are removed from the sample.
\end{enumerate}
To assess the impacts of these effects,
three $\alpha_\mathrm{CO}$ scenarios are considered:
\begin{enumerate}
\item $\alpha_\mathrm{CO}=3.2$, using all pixels (star-forming or not);
\item $\alpha_\mathrm{CO}=3.2$, only using star-forming pixels; and
\item a metallicity-dependent $\alpha_\mathrm{CO}=\alpha_\mathrm{CO}(Z)$ (Eq.~\ref{eq:alphaco_met}).
\end{enumerate}
The impact of only considering star-forming pixels on the total number of 
pixels and galaxies (Table~\ref{tab:selection}) varies depending on how many pixels per galaxy 
are required.
For example, starting from the 95 galaxies in Sample A (Table~\ref{tab:selection}),
if we require at least 4 CO-detected pixels per galaxy, our sample will consist of 83 galaxies
and 2059 pixels (Sample B).
If we require at least 4 CO-detected star-forming pixels per galaxy
(e.g. to apply a metallicity-dependent $\alpha_\mathrm{CO}$),
we would have to remove 43\%\ of the pixels and 22\%\ of the galaxies from the sample, 
and would be left with 
1168 pixels and 64 galaxies (Sample C).
In the analysis that follows, we use Sample C exclusively except for comparison with Sample B in Section~\ref{sec:correlation}.

\section{Analysis and Results}

\subsection{The degree of correlation between $\Sigma(\mathrm{12\>\mu m})$ and $\Sigma(\mathrm{H_2})$}\label{sec:correlation}

\begin{figure*}
	\includegraphics[width=0.35\textwidth]{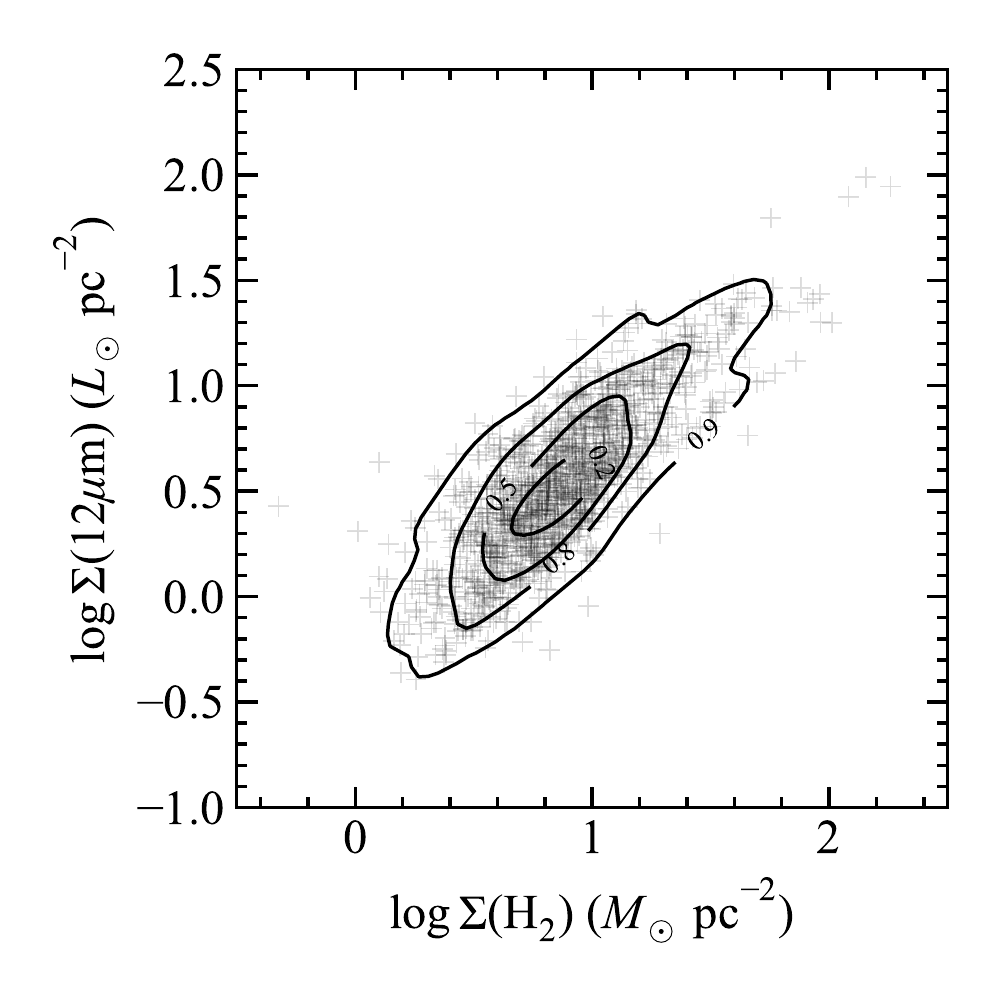}\includegraphics[width=0.35\textwidth]{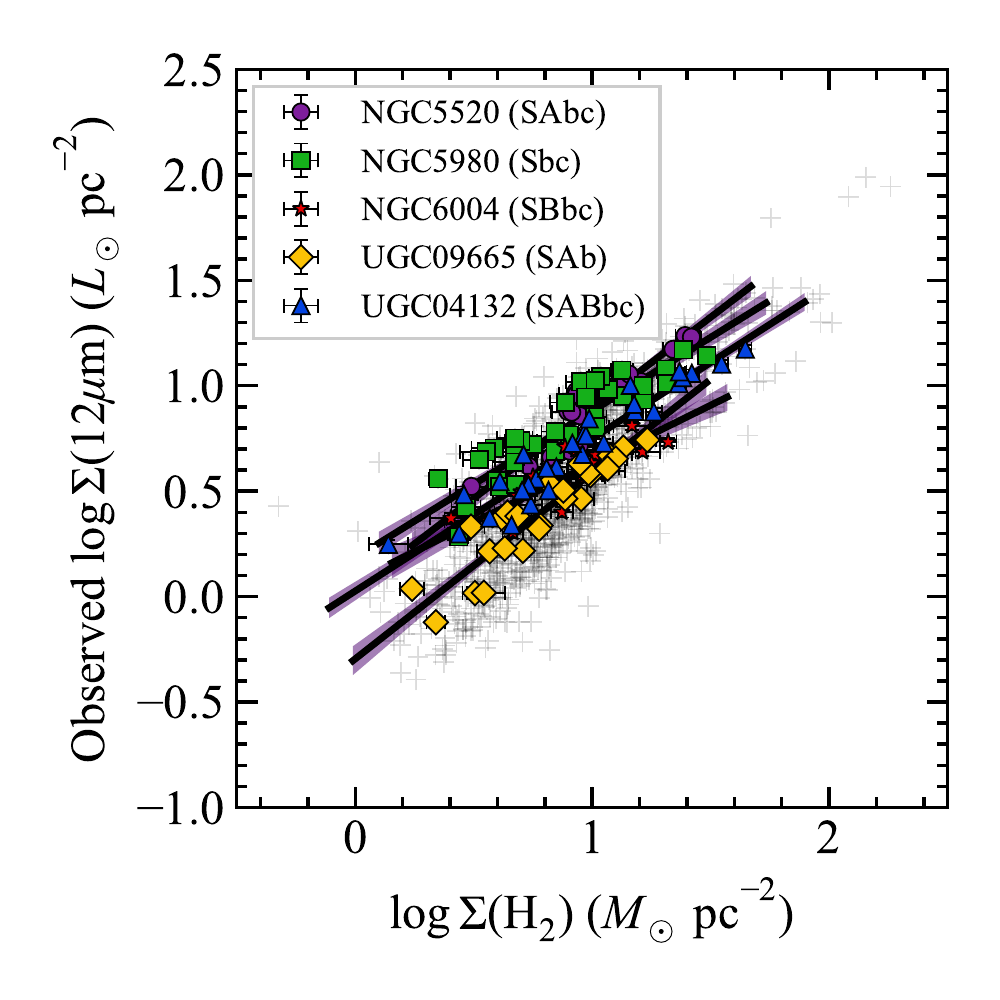}\includegraphics[width=0.35\textwidth]{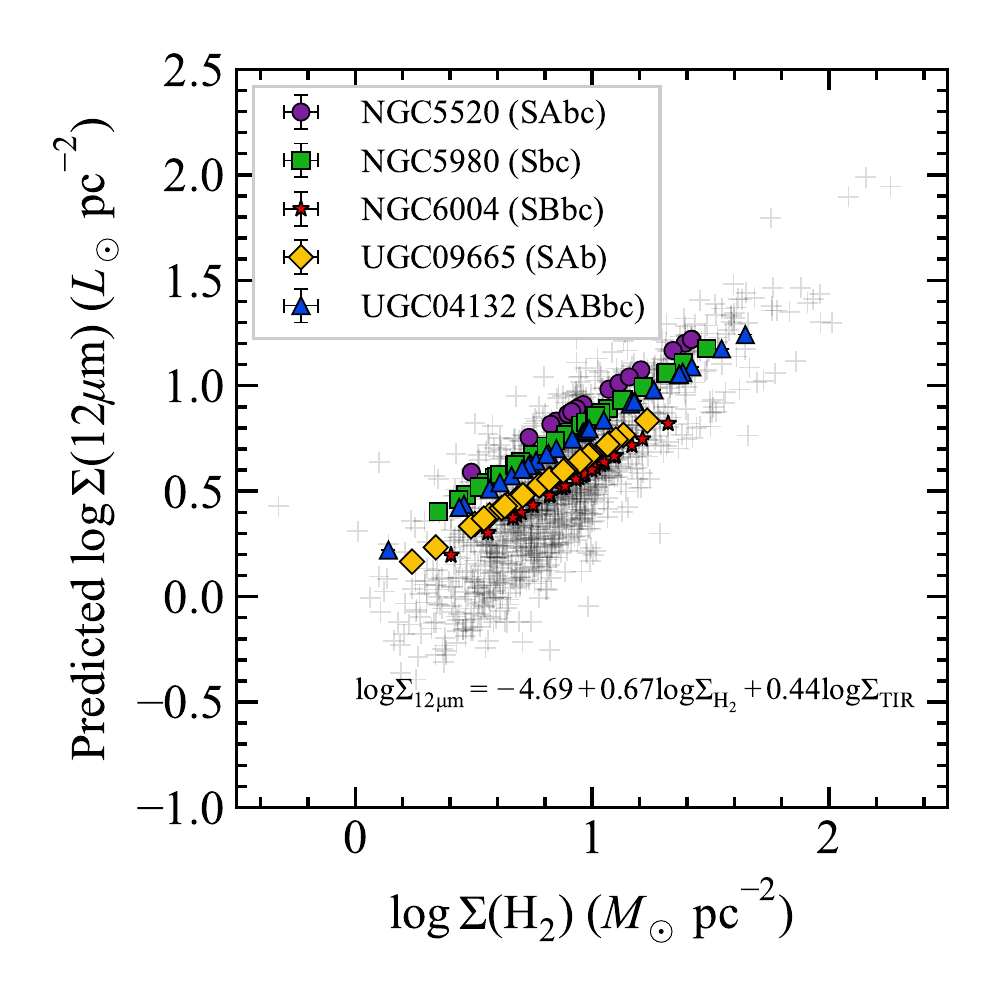}
	\includegraphics[width=0.35\textwidth]{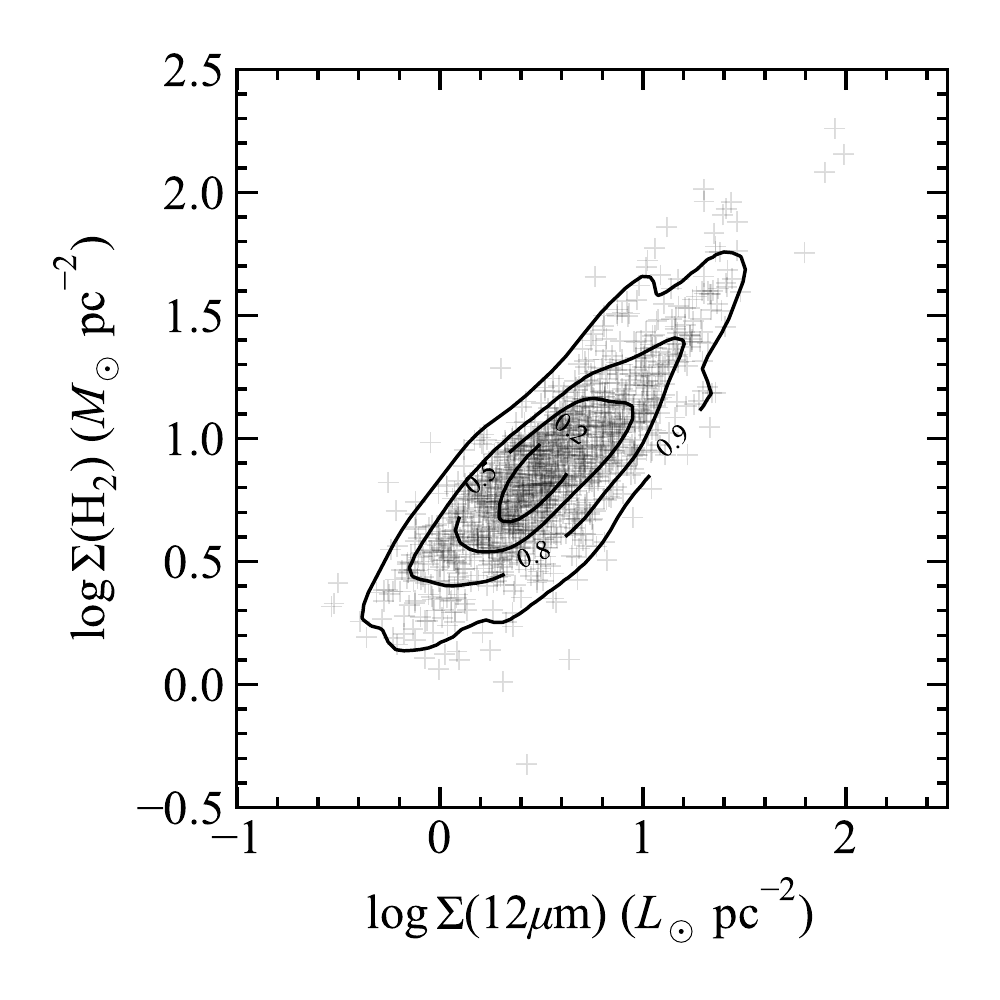}\includegraphics[width=0.35\textwidth]{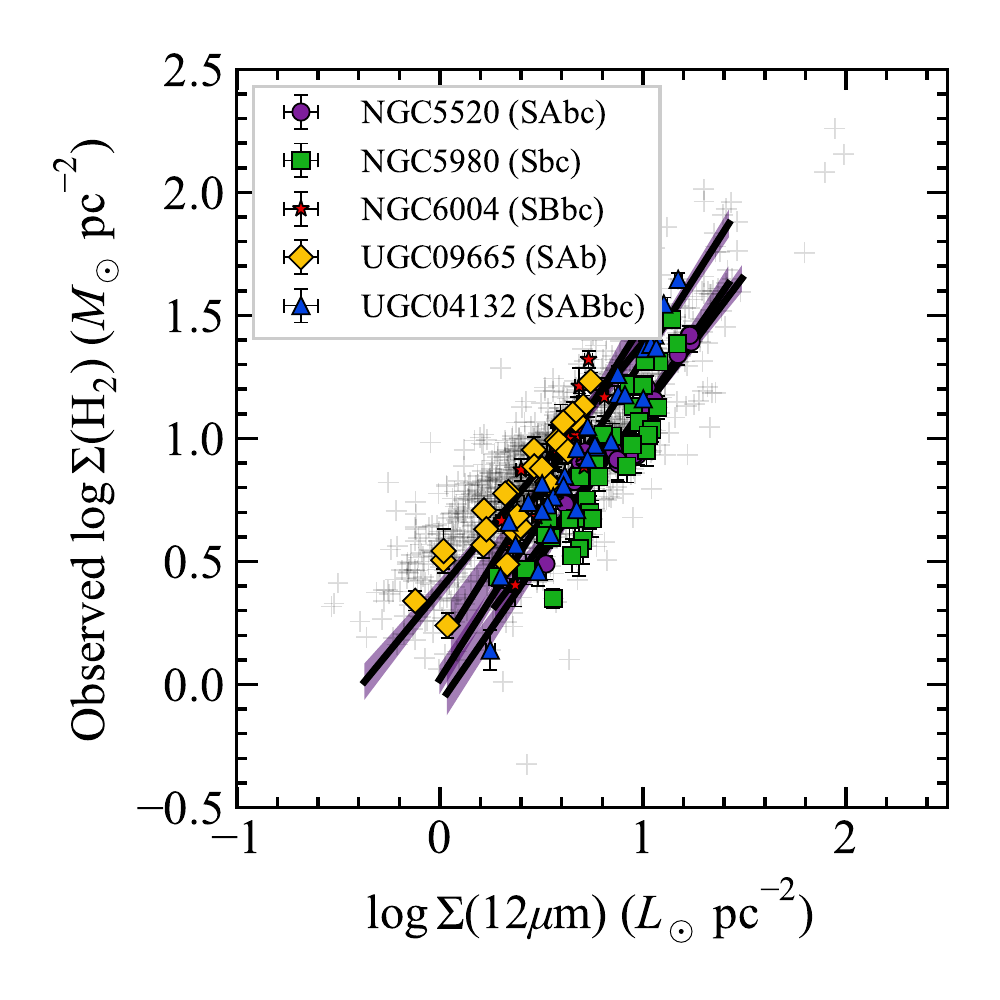}\includegraphics[width=0.35\textwidth]{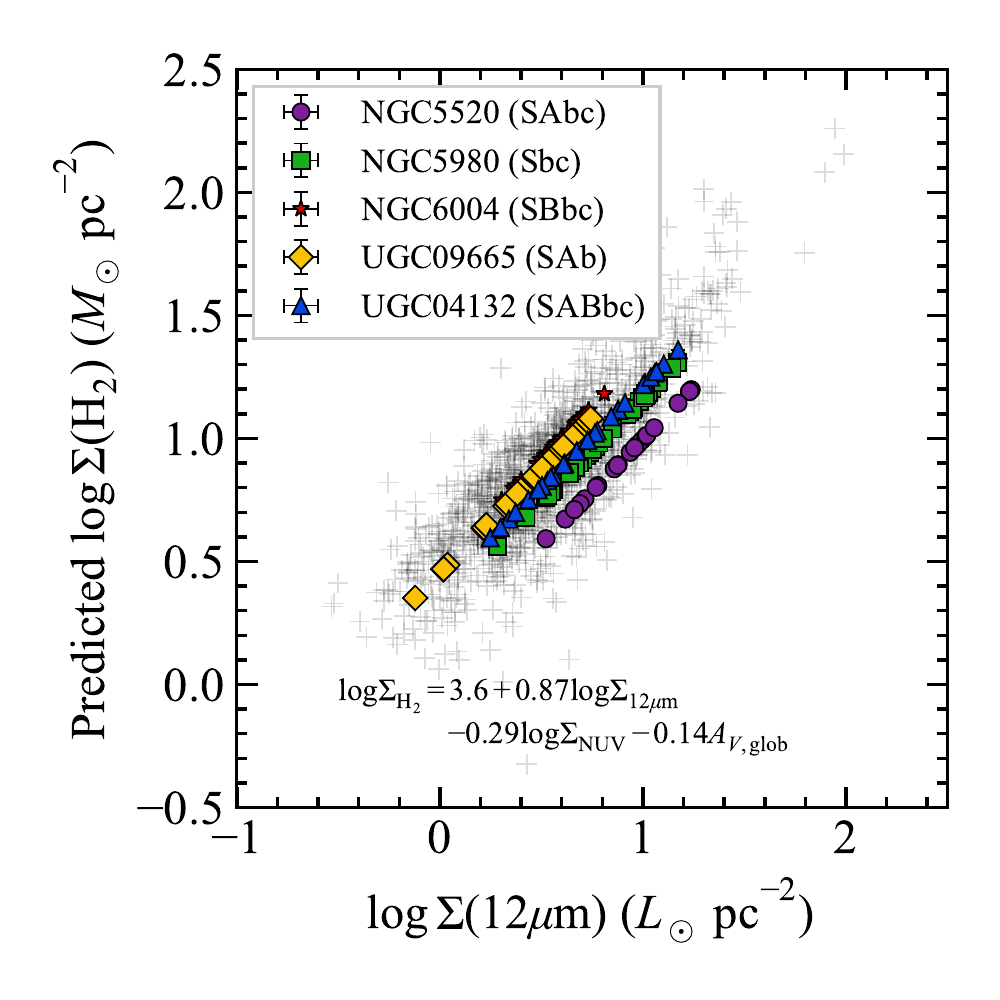}
    \caption{12 \micron\ surface density versus H$_2$ surface density (\textit{top}) and vice versa (\textit{bottom}) assuming a metallicity-dependent $\alpha_\mathrm{CO}$ (Section~\ref{sec:aco}). Only star-forming pixels that are detected in CO are shown.
    \textit{Left:} The grey points are all pixels, and the fraction of pixels 
    enclosed by each contour are indicated. The grey points are the same in all panels. \textit{Middle:} Observed values of $\log\Sigma(\mathrm{12\>\mu m})$ (top) and $\log\Sigma(\mathrm{H_2})$ (bottom) are shown on the y-axes. 
    The pixel values and best linear fits for five example galaxies from Sample C (Table~\ref{tab:selection}) are coloured to illustrate some of the variation in the correlations found. Hubble types from CALIFA DR3 are indicated in the legend for the five selected galaxies. \textit{Right:} Predicted values are shown on the y-axes using selected multi-parameter estimators (Table~\ref{tab:estimators_all}). 
    The predictions were made from fits to the pixels from all galaxies except for the 
    galaxy being predicted, to mimic the case where these estimators would be used on a galaxy outside of the sample in this work.}
    \label{fig:sigma12_sigmah2_overall}
\end{figure*}

Previous work has shown a strong correlation 
between integrated \wise\ 12 \micron\ luminosity and 
CO(1-0) luminosity \citep{jiang2015, gao2019}. 
To determine if this correlation holds at 
sub-galaxy spatial scales, we matched the resolution of 
the EDGE CO maps to \wise\ W3 resolution and compared surface densities 
pixel-by-pixel for each galaxy (Figure~\ref{fig:sigma12_sigmah2_overall}). 
This comparison indicates that there is a clear correlation 
between $\Sigma(\mathrm{12\>\mu m})$ and $\Sigma(\mathrm{H_2})$,
and that within galaxies, the correlation is strong.

\begin{figure*}
	\includegraphics[width=\columnwidth]{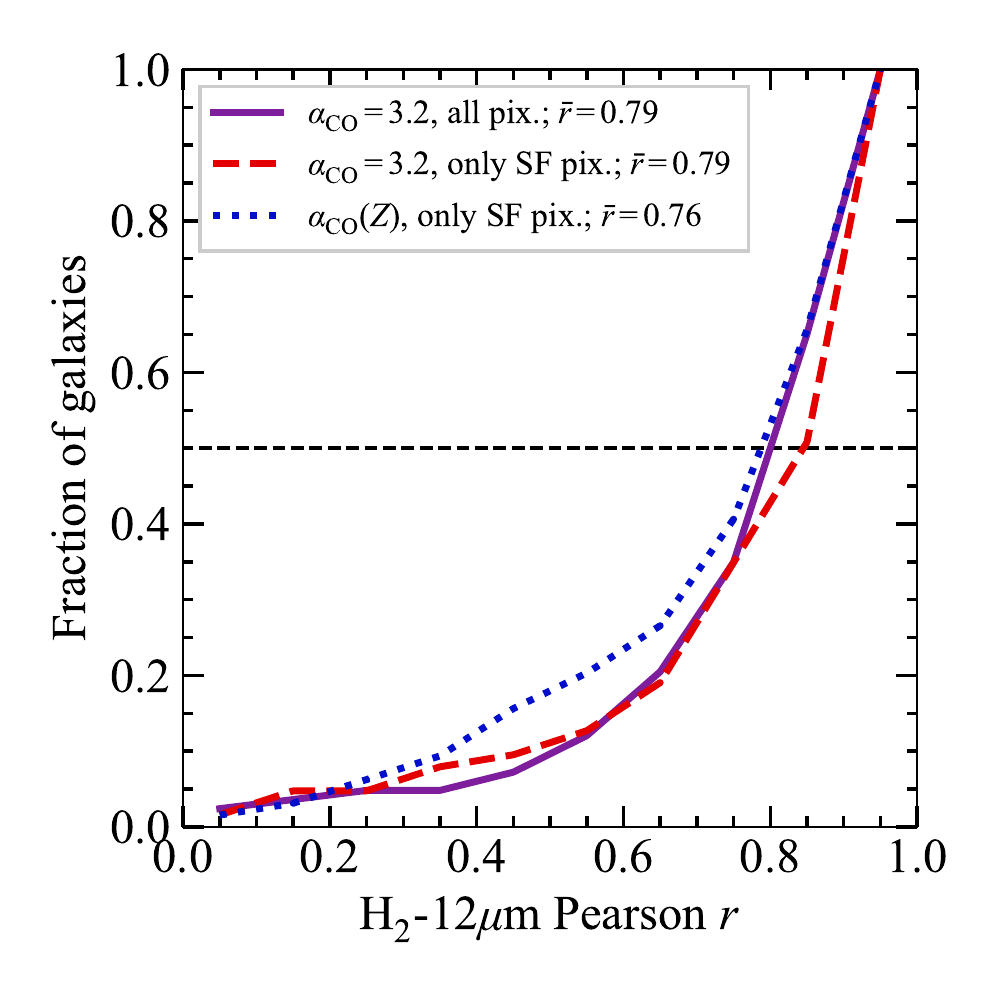}\includegraphics[width=\columnwidth]{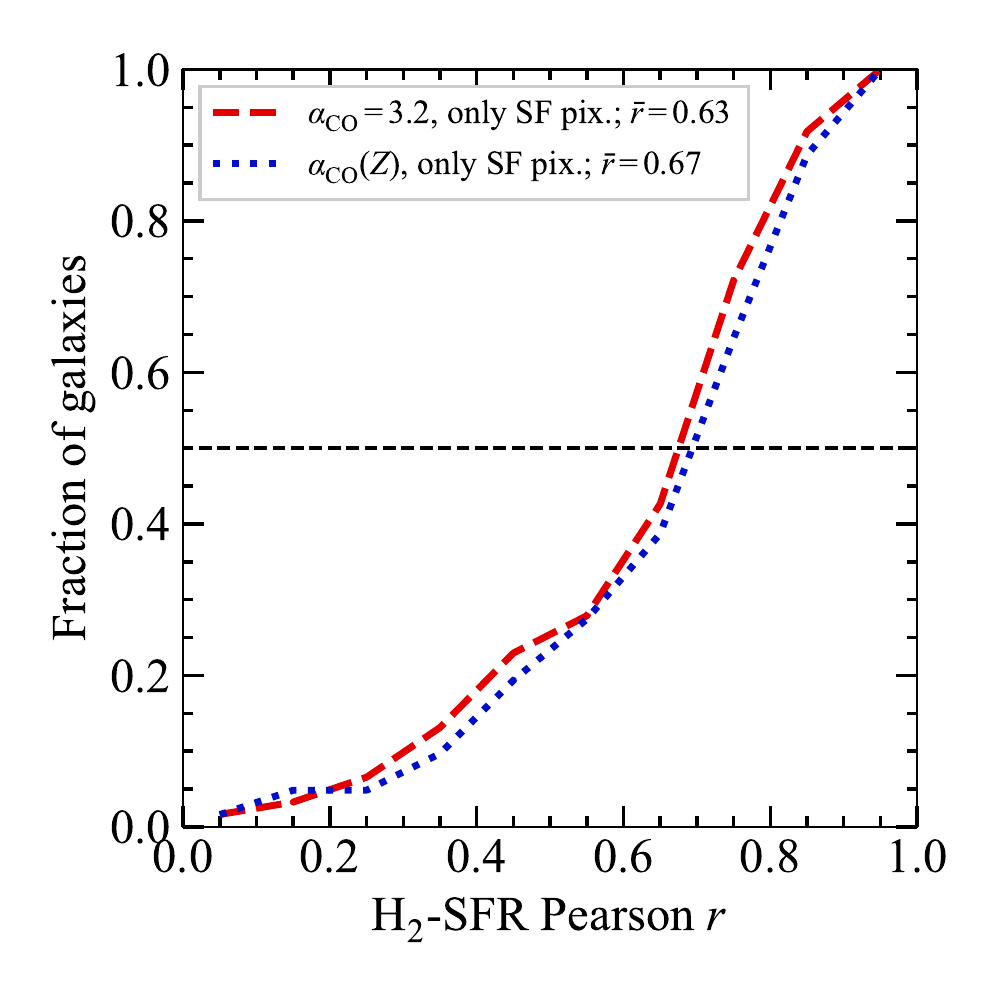}
    \caption{\textit{Left:} Cumulative histogram of the Pearson correlation 
    coefficient between $\log \Sigma(\mathrm{H_2})$ and 
    $\log \Sigma (12\>\micron)$ for each galaxy with a minimum of 4 
    CO-detected pixels each. \textit{Right:} Same as left except between $\log \Sigma(\mathrm{H_2})$ and 
    $\log \Sigma_\mathrm{SFR}$.
    The three colours are for different $\alpha_\mathrm{CO}$ assumptions:
    (1) $\alpha_\mathrm{CO}=3.2$ and including all pixels, (2) 
    $\alpha_\mathrm{CO}=3.2$ including only star-forming pixels,
    and (3) metallicity-dependent $\alpha_\mathrm{CO}$ (Eq.~\ref{eq:alphaco_met}).
    There are
    83, 64, and 64 galaxies shown in the purple, red, and blue histograms respectively. 
    A strong correlation 
    is found for most galaxies, for each $\alpha_\mathrm{CO}$ assumption; however, the mean and
    median correlations between $\Sigma_\mathrm{SFR}$ and $\Sigma(\mathrm{H_2})$ are not as strong as those between $\Sigma(\mathrm{12\>\mu m})$ and $\Sigma(\mathrm{H_2})$. 
    The same galaxies and pixels were used in both panels, so the differences are not due to a selection effect.}
    \label{fig:pearson_hist}
\end{figure*}

To quantify the strength of the correlation per galaxy,
the Pearson correlation coefficient between 
$\log\Sigma(\mathrm{12\>\micron})$ and $\log\Sigma(\mathrm{H_2})$ was calculated 
for each galaxy. 
The distribution of correlation coefficients across all galaxies was computed
separately for each $\alpha_\mathrm{CO}$ scenario (Section~\ref{sec:aco}; Figure~\ref{fig:pearson_hist}).
The means for the three distributions are: 
\begin{enumerate}
\item 0.79 for $\alpha_\mathrm{CO}=3.2$, all pixels included;
\item 0.79 for $\alpha_\mathrm{CO}=3.2$, star-forming pixels only; and
\item 0.76 for $\alpha_\mathrm{CO}(Z)$ (Eq.~\ref{eq:alphaco_met}).
\end{enumerate}
These results indicate that there are 
strong correlations between $\Sigma(\mathrm{12\>\mu m})$ and $\Sigma(\mathrm{H_2})$
regardless of the $\alpha_\mathrm{CO}$ assumed.
A minority of galaxies show poor correlations (4 out of 95 galaxies with correlation coefficients $<0.2$). 
Reasons for poor correlations include fewer CO-detected pixels, and 
small dynamic range in the pixels that are detected (e.g. a region covering multiple pixels 
with uniform surface density).

For comparison, cumulative histograms of the correlation coefficients
between $\log\Sigma_\mathrm{SFR}$ (Eq.~\ref{eq:sigma_sfr}) and $\log\Sigma(\mathrm{H_2})$
were computed (right panel of Figure~\ref{fig:pearson_hist}). The 
same sets of galaxies and pixels were used as in the left panel of Figure~\ref{fig:pearson_hist},
except the ``$\alpha_\mathrm{CO}=3.2$, all pix.'' version is excluded, 
because $\log\Sigma_\mathrm{SFR}$
can only be calculated in star-forming pixels. The mean and median correlation
coefficients are lower than those in the left panel of Figure~\ref{fig:pearson_hist}. Since 
the same pixels are used, this suggests a 
stronger correlation
between $\Sigma(\mathrm{12\>\mu m})$ and $\Sigma(\mathrm{H_2})$ than between $\Sigma_\mathrm{SFR}$ and $\Sigma(\mathrm{H_2})$.

\subsection{Bayesian linear regression}\label{sec:bayesfit}

The relationship between 12 \micron\ and CO emission resembles the 
Kennicutt-Schmidt relation, which also shows variation from galaxy to galaxy 
\citep{shetty2013}.
We model the relationship 
between $\log \Sigma(\mathrm{12\>\micron})$ and
$\log \Sigma(\mathrm{H_2})$ with a power-law
\begin{equation} \label{eq:powerlaw}
\log \Sigma(\mathrm{H_2}) = N\log \Sigma(\mathrm{12\>\micron}) + \log C.
\end{equation}
To determine whether the 12 \micron-CO relation is universal or not, 
we performed linear fits of 
$\log \Sigma (\mathrm{H_2})$ against $\log \Sigma (12\>\micron)$
for each galaxy with at least 4 CO-detected star-forming pixels
(Sample C in Table~\ref{tab:selection}; middle panel of Figure~\ref{fig:sigma12_sigmah2_overall}).
A metallicity-dependent $\alpha_\mathrm{CO}$ was used in Figure~\ref{fig:sigma12_sigmah2_overall}.
These fits were performed using LinMix, a Bayesian linear regression
code which incorporates uncertainties in both $x$ and $y$ \citep{kelly2007}.
We repeated the fits for each $\alpha_\mathrm{CO}$ (Sec.~\ref{sec:aco})
and with 
$\log \Sigma (\mathrm{H_2})$ on the x-axis instead.

\begin{figure*}
	\includegraphics[width=1.75\columnwidth]{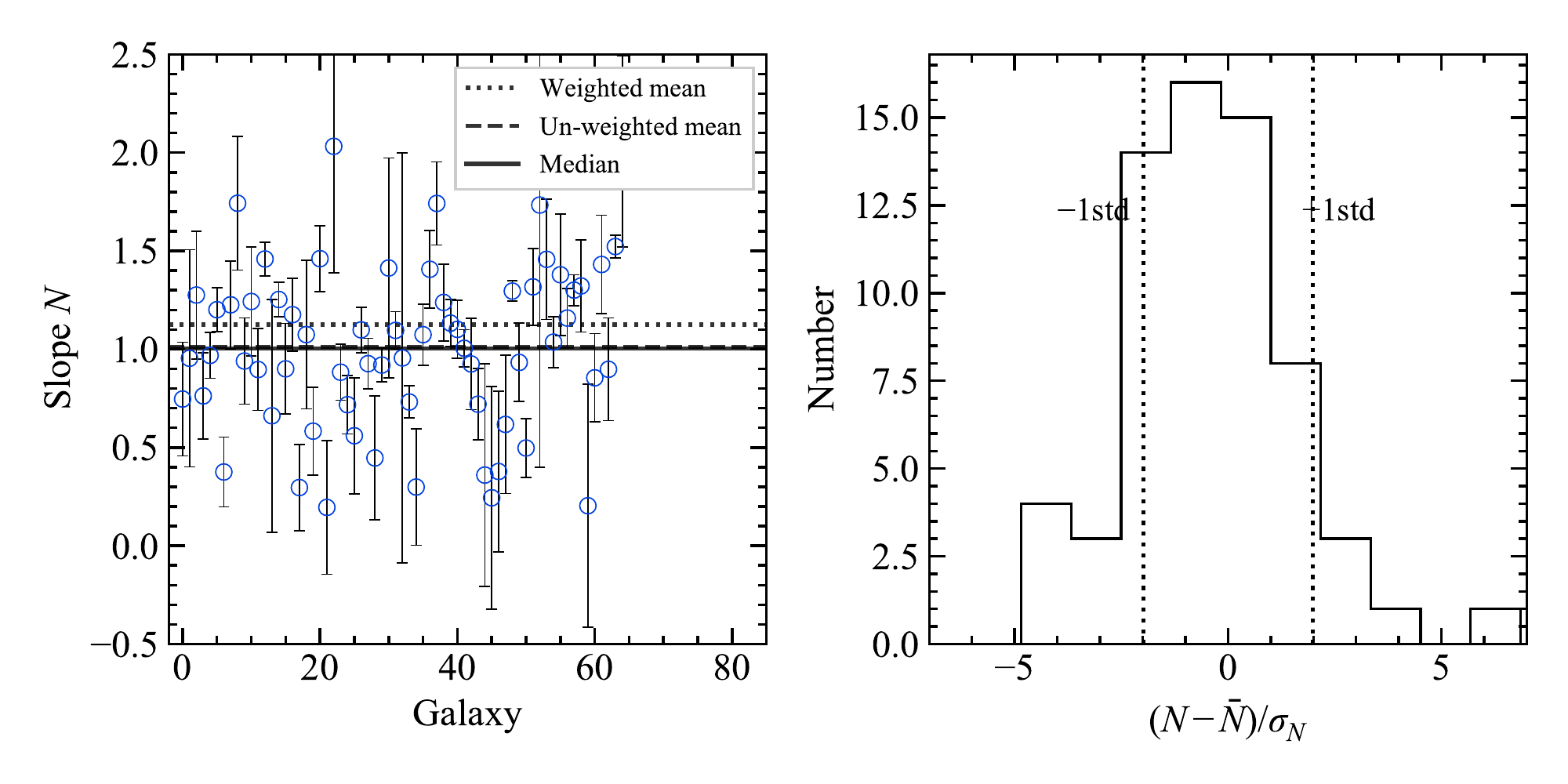}
	\includegraphics[width=1.75\columnwidth]{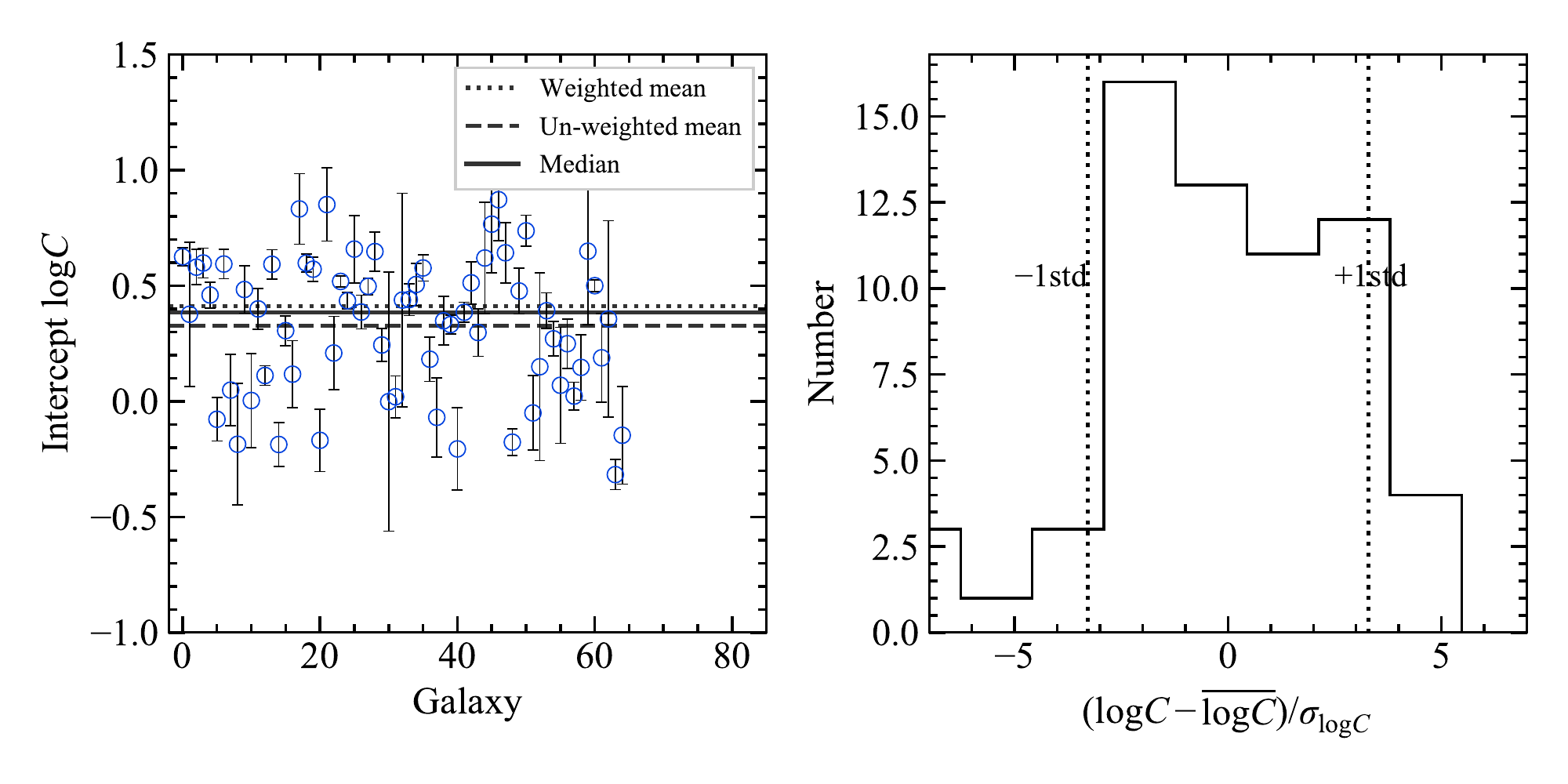}
    \caption{
    Best-fit slope $N$ (top) and intercept $\log C$ (bottom)
    of fits to individual pixel measurements of 
    $\log \Sigma(\mathrm{12\>\mu m})$ (x-axis)
    versus $\log \Sigma(\mathrm{H_2})$ (y-axis).
    Each point is for one galaxy.
    A metallicity-dependent $\alpha_\mathrm{CO}$ was used, 
    so only star-forming pixels were used in the fits.
    At least 4 CO-detected star-forming pixels per galaxy were required 
    (Sample C, Table~\ref{tab:selection}).
    Left: The horizontal lines show the inverse-variance weighted means (dotted),
    un-weighted means (solid), and medians (dashed).
    Right: Histograms of the residuals for each galaxy relative to
    the weighted mean, divided by the uncertainty for each galaxy. The vertical lines indicate 
    $\pm 1$ times the standard deviation of each distribution.
    }
    \label{fig:slope_int_summary}
\end{figure*}

For a given galaxy, the best-fit parameters do not vary much 
depending on the $\alpha_\mathrm{CO}$ assumed, provided 
there are enough pixels to perform the fit even after excluding 
non-starforming pixels.
The fit parameters are also not significantly different if we include
upper limits in the fitting.
However, we find significant 
differences in the slope and intercept
from galaxy to galaxy,
indicating a non-universal resolved relation.
The galaxy-to-galaxy variation in best-fit parameters persists
for all three $\alpha_\mathrm{CO}$ scenarios.
The galaxy-to-galaxy variation can be seen in the distribution of slopes and intercepts 
assuming a metallicity-dependent $\alpha_\mathrm{CO}$ for example (Figure~\ref{fig:slope_int_summary}).
The best-fit intercepts 
span a range of $\simeq1$ dex 
($-0.31$ to $0.87$, median $0.41$),
and the slopes range from 
0.20 to 2.03, with a median of 1.13.
To quantify the significance of the galaxy-to-galaxy variation in best-fit parameters,
residuals in the parameters relative to the mean parameters were computed.
For example, if the measurement of the slope for galaxy $i$ is 
$N_i \pm \sigma_{N_i}$, the residual relative to the average slope over all galaxies 
$\bar{N}$ is $(N_i - \bar{N})/\sigma_{N_i}$.
Similarly, if the measurement of the intercept for galaxy $i$ is 
$\log C_i \pm \sigma_{\log C_i}$, the residual relative to the average intercept over all galaxies 
$\overline{\log C}$ is $(\log C_i - \overline{\log C})/\sigma_{\log C_i}$.
The residual histograms (Figure~\ref{fig:slope_int_summary}) show that
most of the slopes $N_i$ are within $\simeq1.5\sigma_{N_i}$ of $\bar{N}$, 
but the intercepts show more significant deviations (many beyond $3\sigma_{\log C_i}$).

\begin{figure*}
	\includegraphics[width=\columnwidth]{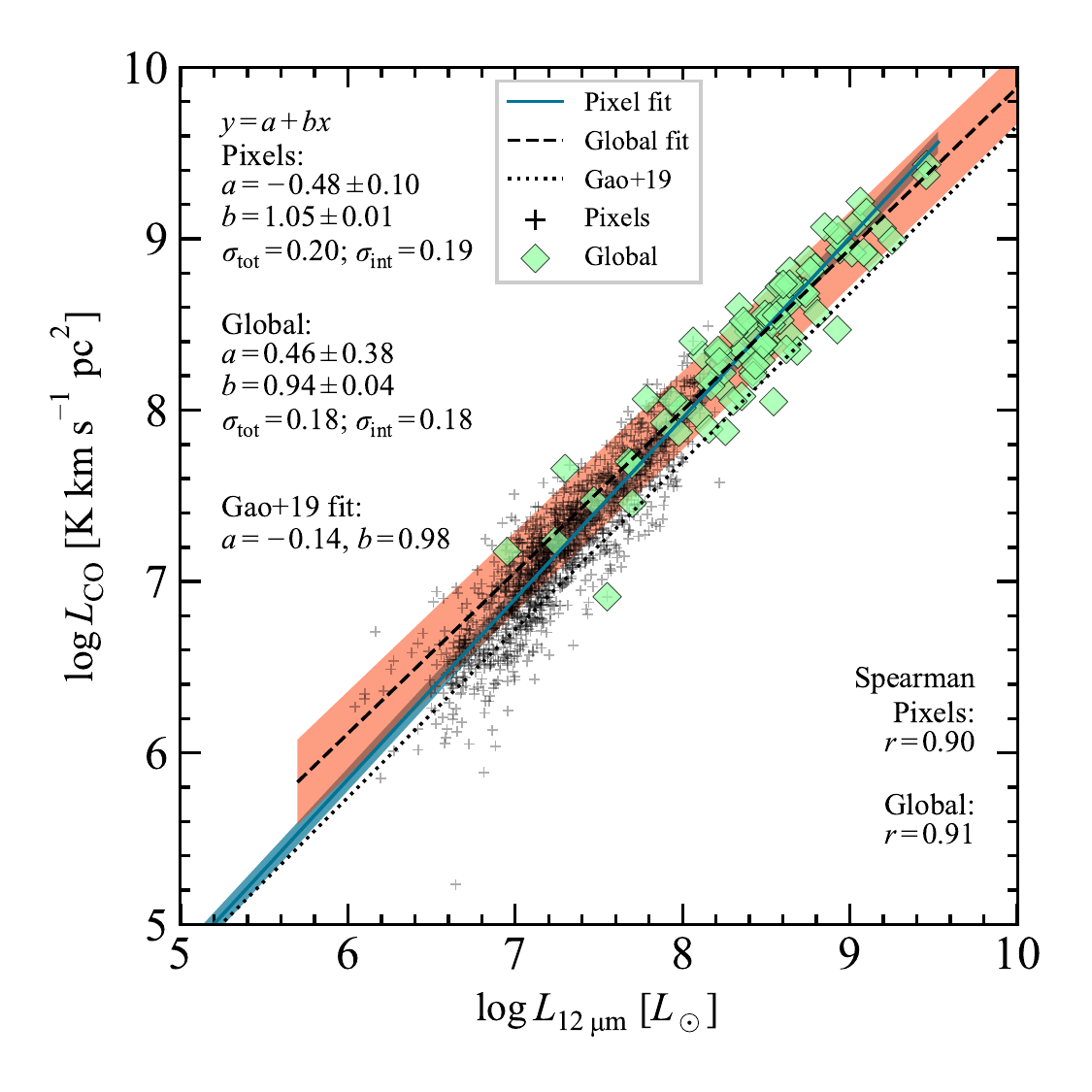}\includegraphics[width=\columnwidth]{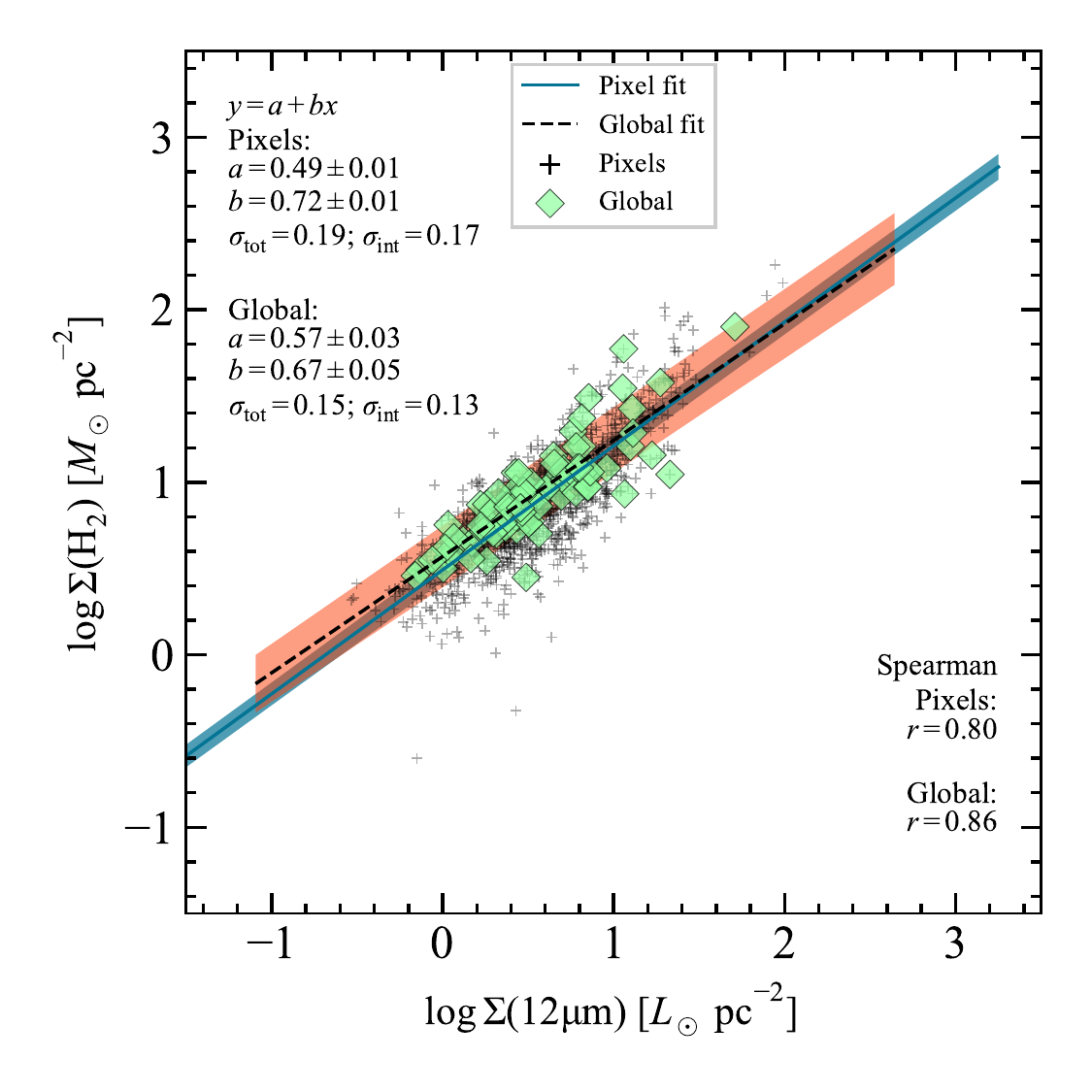}
    \caption{
    Measurements of 12 \micron\ and H$_2$ (or CO) using all 
    individual pixels from all galaxies in the sample (black crosses),
    and the galaxy-integrated values (diamonds).
    The fits (Section~\ref{sec:bayesfit}) were done separately for the pixel measurements (blue regions) and the 
    global measurements (red regions).
    Best-fit parameters assuming a power-law model (Equation~\ref{eq:powerlaw}), and the 
    total $\sigma_\mathrm{tot}$ and intrinsic $\sigma_\mathrm{int}$ scatter (Appendix~\ref{appendix:scatter}) about the fits are indicated.
    The left and right panels show the fits to luminosities and surface densities respectively. H$_2$ surface densities were calculated using a metallicity-dependent $\alpha_\mathrm{CO}$ (Equation~\ref{eq:alphaco_met}). Note that in the right hand panel, ignoring $\alpha_\mathrm{CO}$ uncertainty means that the $\Sigma(\mathrm{H_2})$ uncertainty has been underestimated, and therefore the intrinsic scatter $\sigma_\mathrm{int}$ (derived from $\sigma_\mathrm{tot}$ and the uncertainty on $\Sigma(\mathrm{H_2})$, Equation~\ref{eq:sig_int}) has been overestimated. Also, if we replace $\Sigma(\mathrm{H_2})$ with $\Sigma(\mathrm{CO})$, $\sigma_\mathrm{tot}$ decreases by only 0.01 dex and $\sigma_\mathrm{int}$ does not change, which indicates that the scatter is dominated by that of the 12 \micron-CO surface density relationship. Consequently, $\sigma_\mathrm{int}$ in the right panel should be interpreted as the intrinsic scatter in the 12 \micron-CO surface density relationship.
 For completeness, versions of these plots using the same data but with the x and y axes interchanged are shown in Figure~\ref{fig:w3_vs_co_v3}, and versions with a constant $\alpha_\mathrm{CO}$ and non-starforming pixels included are shown in Figure~\ref{fig:w3_vs_co_v2}.}
    \label{fig:w3_vs_co}
\end{figure*}

To establish how well-fit all pixels are to a single model,
linear fits were done
on all CO-detected pixels from all 83 galaxies in Sample B (Table~\ref{tab:selection}) 
using LinMix (black crosses in Figure~\ref{fig:w3_vs_co}).
The fits were done separately for 
luminosities ($\log L_\mathrm{12\>\mu m}$, $\log L_\mathrm{CO}$; left panel of Figure~\ref{fig:w3_vs_co}) and 
surface densities ($\log \Sigma(\mathrm{12\>\mu m})$, $\log \Sigma(\mathrm{H_2})$; 
right panel of Figure~\ref{fig:w3_vs_co}).
For completeness, the fits were also done with CO/H$_2$ on the x-axis
(Figure~\ref{fig:w3_vs_co_v3}).
In all cases there are strong correlations (correlation coefficients of $\simeq 0.90$), and good fits (total scatter 
about the fit $\sigma_\mathrm{tot} \simeq 0.19$ dex). 
By comparing the total scatter $\sigma_\mathrm{tot}$ and intrinsic scatter $\sigma_\mathrm{int}$ (Appendix~\ref{appendix:scatter}), it is clear that
most of the scatter is intrinsic rather than due to 
measurement and calibration uncertainties.
Note that in the right hand panel of Figure~\ref{fig:w3_vs_co}, ignoring the $\alpha_\mathrm{CO}$ uncertainty means that the $\Sigma(\mathrm{H_2})$ uncertainty has been underestimated, and therefore the intrinsic scatter $\sigma_\mathrm{int}$ (derived from $\sigma_\mathrm{tot}$ and the uncertainty on $\Sigma(\mathrm{H_2})$, Equation~\ref{eq:sig_int}) has been overestimated. Also, if we replace $\Sigma(\mathrm{H_2})$ with $\Sigma(\mathrm{CO})$, $\sigma_\mathrm{tot}$ decreases by only 0.01 dex and $\sigma_\mathrm{int}$ does not change, which indicates that the scatter is dominated by that of the 12 \micron-CO surface density relationship. Consequently, $\sigma_\mathrm{int}$ in the right hand panel of Figure~\ref{fig:w3_vs_co} should be interpreted as the intrinsic scatter in the 12 \micron-CO surface density relationship.

Similarly, to establish how well-fit all \textit{global values} are to a single model,
linear fits were done
on the galaxy-integrated values (green diamonds in Figure~\ref{fig:w3_vs_co}) for all 83 galaxies in Sample B (Table~\ref{tab:selection}).
The results show good fits overall (correlation coefficients of $\simeq 0.90$, scatter 
about the fit $\sigma_\mathrm{tot}\simeq 0.20$ dex).
The global values do indeed follow uniform trends (with the exception of one outlier),
and the global fits with molecular gas on the x-axis show steeper slopes and smaller y-intercepts than the pixel fits (Figure~\ref{fig:w3_vs_co}).
The global fits with 12 \micron\ on the x-axis show shallower slopes and larger y-intercepts than the pixel fits.

\subsection{Spatially resolved estimator of $\Sigma(\mathrm{H_2})$}\label{sec:estimator}

To develop an estimator of $\log \Sigma (\mathrm{H_2})$ from 
$\log \Sigma(\mathrm{12\>\mu m})$ and other galaxy properties,
we performed linear regression on all of the star-forming pixels from all galaxies
combined. Global properties (from UV, optical, and infrared measurements) and resolved optical  properties were included (Table~\ref{tab:fitprops}).
The model is 
\begin{equation}\label{eq:linmodel}
\vec{y} = \theta_0 + \sum_i \theta_i \vec{x_i},
\end{equation}
where each entry of $\vec{y}$ is $\log \Sigma (\mathrm{H_2})$ for each pixel of each galaxy
(using the metallicity-dependent $\alpha_\mathrm{CO}$, Eq.~\ref{eq:alphaco_met}),
the $\theta$ are the fit parameters, and the sum is  
over $i$ properties (a combination of pixel properties or global properties).
We used ridge regression, implemented in the Scikit-Learn Python package \citep{pedregosa2012}, 
which is the same as ordinary least squares regression
except it includes a penalty in the likelihood for more complicated models.
The penalty term is the sum of the squared coefficients of each 
parameter $\delta \sum_i \theta_i^2$.
The regularization parameter $\delta$ (a scalar) sets the impact of the penalty
term. The best value of $\delta$ was determined by cross-validation using \verb'RidgeCV'.
In ridge regression it is important to standardize the data prior to fitting 
(subtract the sample mean and divide by the standard deviation for all global properties and pixel properties)
so that the penalty term is not affected by different units or 
 spreads of the properties.
The standardized version of Equation~\ref{eq:linmodel} is 
\begin{equation}\label{eq:linmodel_scaled}
\vec{y} - \mathrm{mean}(\vec{y}) = \sum_i \tilde{\theta_i} \left[\frac{\vec{x_i} - \mathrm{mean}(\vec{x_i})}{\mathrm{std}(\vec{x_i})}\right].
\end{equation}
Note that it is not necessary to divide $\vec{y} - \mathrm{mean}(\vec{y})$ by 
$\mathrm{std}(\vec{y})$
because it does not impact the regularization term.
After performing ridge regression on the standardized data 
(which provides $\tilde{\theta_i}$), the best-fit coefficients 
in the original units are given by
\begin{equation}\label{eq:linmodel_coeff}
\theta_i = \frac{\tilde{\theta_i}}{\mathrm{std}(\vec{x_i})}.
\end{equation}
The intercept $\theta_0$ is given by
\begin{equation}\label{eq:linmodel_int}
\theta_0 = \mathrm{mean}(\vec{y}) - \sum_i \tilde{\theta_i} \left[\frac{\mathrm{mean}(\vec{x_i})}{\mathrm{std}(\vec{x_i})}\right].
\end{equation}

\begin{table*}
	\centering
	\caption{Global properties (top) and pixel properties (bottom) considered in the multi-parameter fits (Section~\ref{sec:estimator}). The SFR and stellar masses from 
	~\citetalias{bolatto2017} were both multiplied by 0.66 to convert 
	from Salpeter to Kroupa IMF \citep{madau2014}.
	Global SFR, $M_*$, and luminosities were converted to surface densities
	by dividing by $2\pi r_{50}^2$, where $r_{50}$ is the \textit{i}-band 
	half-light radius in kpc from \citet{gilhuly2018}.
	}
	\label{tab:fitprops}
	\begin{tabular}{llll} %
		\hline
		Label & Units & Reference & Description \\
		\hline
		\multicolumn{4}{c}{Global Properties} \\
		\hline
		$12+\log\mathrm{O/H}_\mathrm{glob}$ & dex & \citetalias{bolatto2017} & [\ion{O}{III}]/[\ion{N}{II}]-based gas-phase metallicity \\
		$\log \Sigma_\mathrm{SFR,glob}$ & $\mathrm{M_\odot}$ yr$^{-1}$ kpc$^{-2}$ & \citetalias{bolatto2017} & Star formation rate surface density ($5.3\times 10^{-42}L(\mathrm{H\alpha})/2\pi r_{50}^2$)\\
		$\log \Sigma_\mathrm{*,glob}$ & $\mathrm{M_\odot}$ kpc$^{-2}$ & \citetalias{bolatto2017} & Stellar mass surface density assuming a Kroupa IMF \\
		$\log \cos i$ & & \citetalias{bolatto2017} & Inclination $i$ is either from CO kinematics, H$\alpha$ kinematics, or LEDA \\
		$\log \Sigma_\mathrm{NUV}$ & 10$^{42}$ erg s$^{-1}$ kpc$^{-2}$ &  \citetalias{catalan-torrecilla2015} & Near-UV surface density \\
		$\log \Sigma_\mathrm{FUV}$ & 10$^{42}$ erg s$^{-1}$ kpc$^{-2}$ &  \citetalias{catalan-torrecilla2015} & Far-UV surface density \\
		$\log \Sigma_\mathrm{TIR}$ & 10$^{43}$ erg s$^{-1}$ kpc$^{-2}$ &  \citetalias{catalan-torrecilla2015} & Total-IR (8-1000\micron) surface density \\
		$\log \Sigma_\mathrm{W4}$ & 10$^{42}$ erg s$^{-1}$ kpc$^{-2}$ & \citetalias{catalan-torrecilla2015} & \wise\ W4 (22 \micron) surface density \\
		$u-r$ & mag &  \citetalias{bolatto2017} & Colour from CALIFA synthetic photometry (SDSS filters applied to extinction-corrected spectra) \\
		$b/a$ & & \citetalias{catalan-torrecilla2015} & Minor-to-major axis ratio from CALIFA synthetic photometry \\
		$(B/T)_g$ &  & \citetalias{catalan-torrecilla2015} & Bulge-to-total ratio from $g$-band photometry \\
		$n_g$ & & \citetalias{catalan-torrecilla2015} & S\'ersic index from $g$-band photometry \\
		$\log \sigma_\mathrm{bulge}$ & km s$^{-1}$ & \citetalias{gilhuly2019} & Bulge velocity dispersion (5 arcsec aperture) \\
		$A_{V,\mathrm{glob}}$ & mag & \citetalias{catalan-torrecilla2015} & Extinction measured from the Balmer decrement \\
\hline
		\multicolumn{4}{c}{Pixel Properties} \\
		\hline
		$12+\log\mathrm{O/H}_\mathrm{pix}$ & dex & Eq.~\ref{eq:metallicity} & [\ion{O}{III}]/[\ion{N}{II}]-based gas-phase metallicity \\
		$\log \Sigma_\mathrm{SFR,pix}$ & $\mathrm{M_\odot}$ yr$^{-1}$ kpc$^{-2}$ & Eq.~\ref{eq:sigma_sfr} & Star formation rate surface density\\
		$\log \Sigma_\mathrm{*,pix}$ & $\mathrm{M_\odot}$ pc$^{-2}$ & Sec.~\ref{sec:califa} & Stellar mass surface density, assuming a Kroupa IMF \\
		$A_{V,\mathrm{pix}}$ & mag & Eq.~\ref{eq:a_ha} & Extinction measured from the Balmer decrement \\
		\hline
	\end{tabular}
\end{table*}

Our goal was to identify a combination of 
properties such that the linear fit of $\log\Sigma(\mathrm{H_2})$
vs. these properties (including $\log\Sigma(\mathrm{12\>\mu m})$) was able to reliably predict $\log\Sigma(\mathrm{H_2})$.
The $\log\Sigma(\mathrm{H_2})$-predicting ability of the fit to a given parameter combination
was quantified by performing fits with one galaxy excluded, and then 
measuring the 
mean-square (MS) error of the prediction for the excluded galaxy (the ``testing error'')\begin{equation}\label{eq:mse}
\mathrm{MS\> error} = \frac{1}{N_\mathrm{pix}}\sum_{N_\mathrm{pix}} (y_\mathrm{true}-y_\mathrm{pred})^2,
\end{equation}
where $N_\mathrm{pix}$ is the number of pixels for this galaxy,
$y_\mathrm{true}$ is the true value of $\log \Sigma(\mathrm{H_2})$ in each pixel,
and $y_\mathrm{pred}$ is the predicted value at that pixel using the fit.
The RMS error over all test galaxies
\begin{equation}\label{eq:rmse}
\mathrm{RMS\> error} = \sqrt{\frac{1}{N_\mathrm{galaxies}} \sum_\mathrm{galaxy} \mathrm{MS\> error}_\mathrm{galaxy}}
\end{equation}
was used to 
decide on a best parameter combination. %

To identify the best possible combination of parameters we did the fit separately for all possible combinations with at least one resolved property required in each combination.
We did not want to exclude the possibility of parameters other than 12 \micron\ 
being better predictors of H$_2$, so we included all combinations even if 12 \micron\ was
excluded.
To avoid overfitting, we excluded galaxies if
the number of CO-detected star-forming pixels 
minus the number of galaxy properties in the estimator was less than 4 (so there are at least 3 degrees of freedom per galaxy after doing the fit),
and only considered models with less than 6 independent variables.
We used the metallicity-dependent $\alpha_\mathrm{CO}$, so the 
sample used for these fits was Sample C (Table~\ref{tab:selection});
however, depending on the number of galaxy properties used and the number of CO-detected star-forming pixels, the sample is smaller for some estimators. We require a minimum of 
15 galaxies for each estimator.

\begin{figure}
	\includegraphics[width=0.5\textwidth]{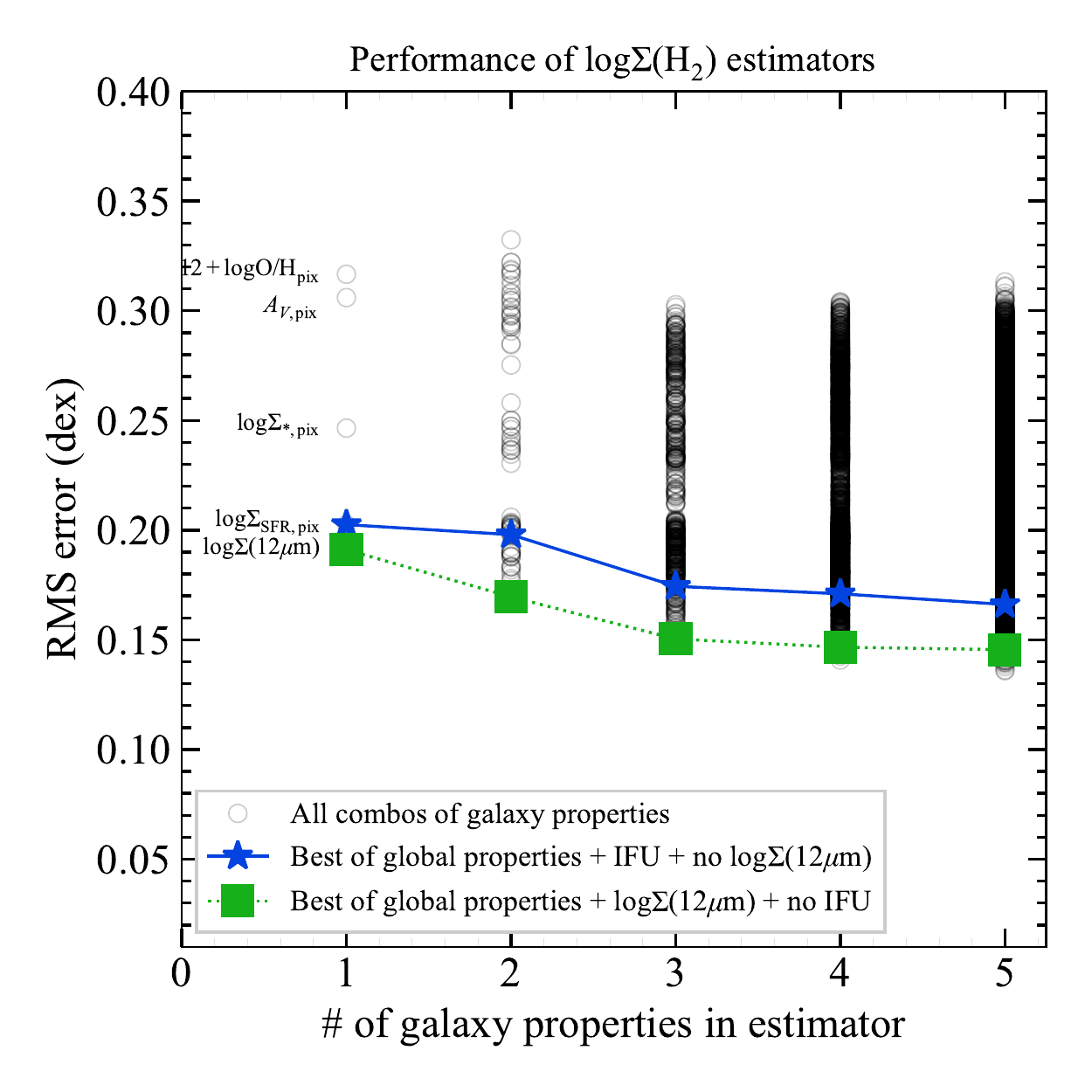}
    \caption{RMS error (Equation~\ref{eq:rmse}) of all estimators. Estimators with smaller RMS errors have better predictive accuracy. The RMS error decreases only slightly as the number of independent variables
    increases for the fits with resolved 12 \micron\ but no IFU data. The fits with resolved 12 \micron\ but no IFU data have lower RMS errors than those with IFU data. The lack of points below the green curve indicates that there is little to be gained by adding IFU data to the estimators with resolved 12 \micron. The RMS error of the estimator with only resolved $A_V$ for example (black circle, upper left) performs significantly worse than the fit with only 12 \micron\ (green square, lower left).}
    \label{fig:rmse}
\end{figure}

Here we describe how the pixel selection and fitting method were used to calculate
the RMS error for each combination of galaxy properties:
\begin{enumerate}
    \item Generate all possible sets of pixels such that each set has the pixels from one galaxy left out.
    \item For each set of pixels:
    \begin{enumerate}
    	\item Compute $\mathrm{mean}(\vec{x_i})$ and $\mathrm{std}(\vec{x_i})$ of the resolved and global properties $\vec{x_i}$. Use these to standardize the data.
    	\item Perform the multi-parameter fit on the standardized data, which yields $\tilde{\theta_i}$ (Eq.~\ref{eq:linmodel_scaled}).
		\item Compute the un-standardized coefficients $\theta_i$ (Eq.~\ref{eq:linmodel_coeff}) and zero-point $\theta_0$ (Eq.~\ref{eq:linmodel_int}).
			\item Use these $\theta_0$, $\theta_i$ to predict $\vec{y}$ of the excluded galaxy (Eq.~\ref{eq:linmodel}).
			\item Tabulate the mean squared-error (Eq.~\ref{eq:mse}).
	\end{enumerate}
		\item Compute the RMS error (Eq.~\ref{eq:rmse}) from all of the MS errors. This indicates the ability of this multi-parameter fit to predict new $\vec{y}$. The RMS error for each estimator is shown in Figure~\ref{fig:rmse}.
\end{enumerate}

In practical applications outside of this work, 
not all of the global properties and pixel 
properties will be available. 
For this reason, we provide several $\log \Sigma(\mathrm{H_2})$ 
estimators which can be used depending on which data are available.
To highlight the relative importance of resolved optical properties vs.
12 \micron, the best-performing estimators based on the following galaxy properties  
are compared:
\begin{enumerate}
    \item all global properties + IFU properties + 12 \micron\ (Table~\ref{tab:estimators_all}),
    \item all global properties + 12 \micron\ but no IFU properties (Table~\ref{tab:estimators_no_ifu}),
    \item all global properties + IFU properties but no 12 \micron\ (Table~\ref{tab:estimators_no_12um}). 
\end{enumerate}
The performance of the estimators was ranked based on their RMS error of 
predicted $\log \Sigma(\mathrm{H_2})$ (Figure~\ref{fig:rmse}). 
The reported estimators are those with the lowest RMS error 
at a given number of galaxy properties (those corresponding to the stars and squares in Figure~\ref{fig:rmse}).
We estimated the uncertainty on the 
coefficients in each estimator by perturbing the 12 \micron\ and H$_2$ data points randomly 
according to their uncertainties, redoing the fits 1000 times, and measuring the 
standard deviation of the parameter distributions. 

The lack of points below the green curve in Figure~\ref{fig:rmse} indicates that there is 
little to be gained by adding IFU data to the estimators with resolved 12 \micron\ (little to no drop in RMS error). 
The RMS error of the estimator with only resolved $A_V$ for example 
(black circle, upper left) performs significantly worse than the
fit with only 12 \micron\ (green square, lower left). 
Estimators with resolved 12 \micron\ but no IFU data perform
better than those with IFU data but no resolved 12 \micron.
There is also no improvement in predictive accuracy of the estimators using global properties + resolved 12 \micron\ + no IFU data beyond a four-parameter fit (intercept, $\Sigma(12\mathrm{\mu m})$, $\Sigma_\mathrm{NUV}$, and global $A_V$).
The best H$_2$ estimators all contain $\log \Sigma (\mathrm{12\>\micron})$,
which indicates that this variable is indeed the most important
for predicting H$_2$.

For the fits in the opposite direction, $\log \Sigma (\mathrm{H_2})$ was found to be the most important for predicting $\mathrm{12\>\micron}$.
The best estimators for 1-5 galaxy properties
show that if $\log \Sigma (\mathrm{H_2})$ is already included, there is essentially no improvement in predictive accuracy (little to no drop in RMS error)
when resolved optical IFU data are included as variables in the fitting.

\begin{figure}
	\includegraphics[width=0.5\textwidth]{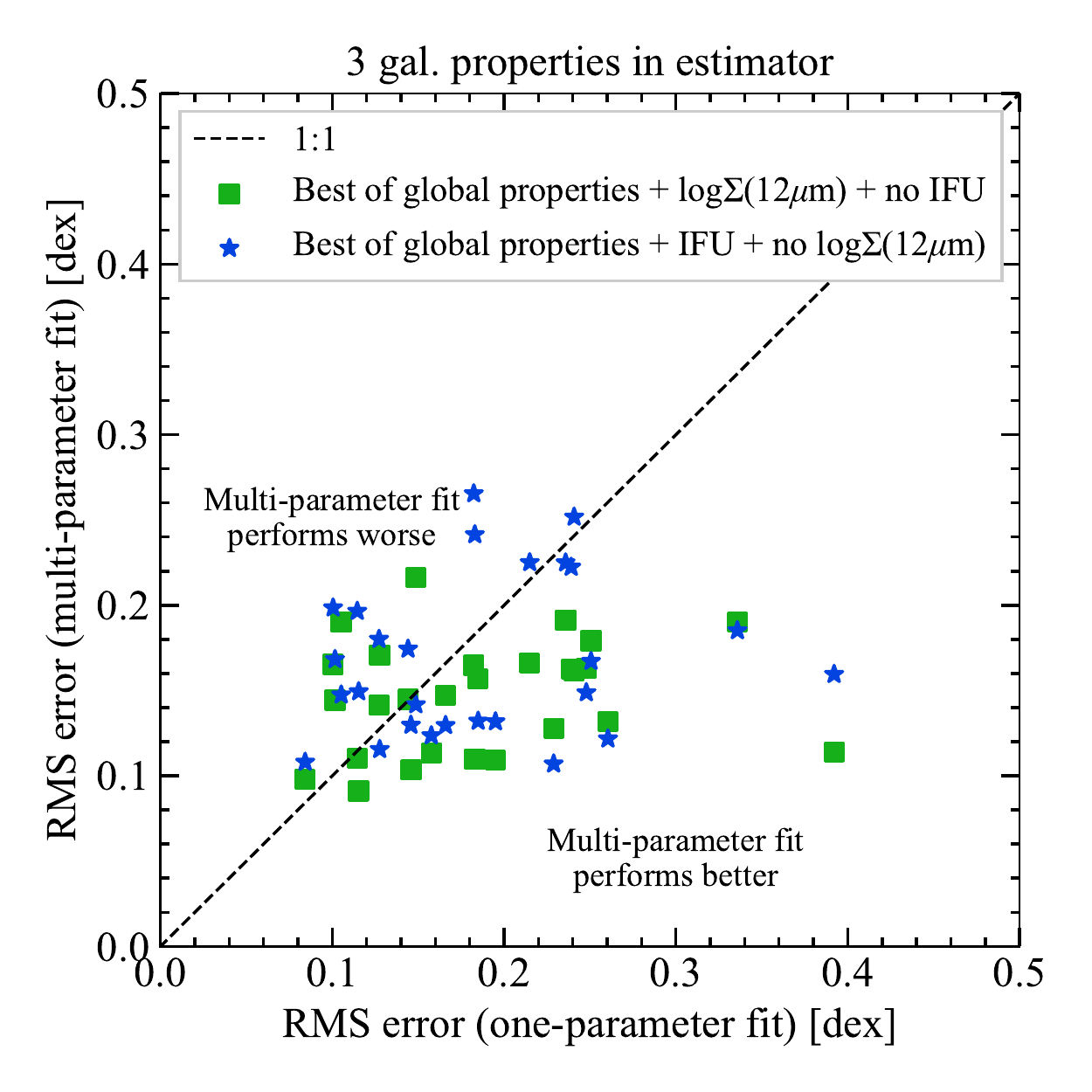}
    \caption{Galaxy-by-galaxy RMS error (Equation~\ref{eq:rmse}) computed from the specified multi-parameter fits with 3 galaxy properties, versus the RMS error computed from the one parameter surface density fit (Figure~\ref{fig:w3_vs_co}). The green squares and blue stars correspond to the green square and blue star in Figure~\ref{fig:rmse} at $n=3$ respectively. The RMS of the y-values of the green squares here gives the RMS error at $n=3$ in Figure~\ref{fig:rmse}, and likewise for the blue stars (Equation~\ref{eq:rmse}).}
    \label{fig:rmse_2}
\end{figure}

We compared how well these multi-parameter estimators perform relative to the
one-parameter estimator 
from the right panel of Figure~\ref{fig:w3_vs_co}:
\begin{equation}
\log \Sigma(\mathrm{H_2}) = (0.49 \pm 0.01) + (0.72 \pm 0.01) \log\Sigma(12\mathrm{\mu m}).
\end{equation}
Note that this fit, obtained via Bayesian linear regression (Sec.~\ref{sec:bayesfit}) 
is consistent with the result from ridge regression (first row of Table~\ref{tab:estimators_all}).
To compare the performance of each estimator with the fit above, 
predicted $\log \Sigma(\mathrm{H_2})$ 
for each pixel was computed from the one-parameter fit,
and the RMS error (square root of Eq.~\ref{eq:mse}) was computed for each galaxy (Figure~\ref{fig:rmse_2}).
Most points lie below 
the 1:1 relation in Figure~\ref{fig:rmse_2}, indicating that
the multi-parameter fits have lower RMS error per pixel than 
the single-parameter fit.

\begin{table*}
	\centering
	\caption{Best-performing estimators of $\log\Sigma(\mathrm{H_2})$ (metallicity-dependent $\alpha_\mathrm{CO}$, Sec.~\ref{sec:aco}) 
	based on global properties + resolved 12 \micron\ + resolved optical IFU properties (Table~\ref{tab:fitprops}). 
	Each successive row adds one galaxy property. For example, the estimator in the second row is  $\log\Sigma(\mathrm{H_2}) = 2.54 + 0.78 \log \Sigma(\mathrm{12\>\mu m}) - 0.20 \log \Sigma_\mathrm{FUV}$. The RMS error (the accuracy of predicted 
	$\log\Sigma(\mathrm{H_2})$ per pixel, Eq.~\ref{eq:rmse}), the number of galaxies $n_\mathrm{gal}$ and pixels $n_\mathrm{pix}$ used for the fit, and the
	intrinsic scatter ($\sigma_\mathrm{int}$, Appendix~\ref{appendix:scatter}) are reported. Table~\ref{tab:estimators_all_aco3p2} shows the best-fit results assuming $\alpha_\mathrm{CO} =3.2$.}
	\label{tab:estimators_all}
	\begin{tabular}{lllll|cc|ccc} %
		\hline
		& & & & & \multicolumn{2}{c|}{$\theta_i$ for pixel properties} & \multicolumn{3}{c|}{$\theta_i$ for global properties} \\
		\cmidrule(lr){6-7}  \cmidrule(lr){8-10} 
		RMS error & $n_\mathrm{gal}$ & $n_\mathrm{pix}$ & $\sigma_\mathrm{int}$ & Zero-point ($\theta_0$) & $\log \Sigma(\mathrm{12\>\mu m})$ & $(12+\log \mathrm{O/H})$ & $\log \Sigma_\mathrm{FUV}$ & $\log \Sigma_\mathrm{NUV}$ & $A(\mathrm{H\alpha})$  \\
		\hline
  $0.19$ &  $58$ &  $1126$ &  $0.17$ & $0.48 \pm 0.01$ & $0.71 \pm 0.01$ &  -- &  -- &  -- &  --  \\
  $0.17$ &  $30$ &  $573$ &  $0.15$ & $2.54 \pm 0.07$ & $0.78 \pm 0.01$ &  -- & $-0.20 \pm 0.01$ &  -- &  --  \\
  $0.15$ &  $27$ &  $552$ &  $0.15$ & $3.6 \pm 0.1$ & $0.87 \pm 0.01$ &  -- &  -- & $-0.29 \pm 0.01$ & $-0.14 \pm 0.01$  \\
  $0.14$ &  $27$ &  $552$ &  $0.14$ & $14.6 \pm 0.6$ & $0.94 \pm 0.01$ & $-1.24 \pm 0.07$ &  -- & $-0.30 \pm 0.01$ & $-0.15 \pm 0.01$  \\
  		\hline
	\end{tabular}
\end{table*}

\begin{table*}
	\centering
	\caption{Same as Table~\ref{tab:estimators_all} but the best-performing estimators based on global properties + resolved 12 \micron\  but no resolved optical IFU properties. Table~\ref{tab:estimators_no_ifu_aco3p2} shows the best-fit results assuming $\alpha_\mathrm{CO} = 3.2$.}
	\label{tab:estimators_no_ifu}
	\begin{tabular}{lllll|c|ccc} %
		\hline
		& & & & & $\theta_i$ for pixel properties & \multicolumn{3}{c|}{$\theta_i$ for global properties} \\
		\cmidrule(lr){6-6}  \cmidrule(lr){7-9} 
		RMS error & $n_\mathrm{gal}$ & $n_\mathrm{pix}$ & $\sigma_\mathrm{int}$ & Zero-point ($\theta_0$) & $\log \Sigma(\mathrm{12\>\mu m})$ & $\log \Sigma_\mathrm{FUV}$ & $\log \Sigma_\mathrm{NUV}$ & $A(\mathrm{H\alpha})$ \\
		\hline
  $0.19$ &  $58$ &  $1126$ &  $0.17$ & $0.47 \pm 0.01$ & $0.71 \pm 0.01$ &  -- &  -- &  --  \\
  $0.17$ &  $30$ &  $573$ &  $0.15$ & $2.54 \pm 0.07$ & $0.78 \pm 0.01$ & $-0.20 \pm 0.01$ &  -- &  --  \\
  $0.15$ &  $27$ &  $552$ &  $0.15$ & $3.6 \pm 0.1$ & $0.88 \pm 0.01$ &  -- & $-0.29 \pm 0.01$ & $-0.14 \pm 0.01$  \\
  $0.15$ &  $27$ &  $552$ &  $0.15$ & $3.6 \pm 0.1$ & $0.87 \pm 0.01$ & $0.03 \pm 0.03$ & $-0.31 \pm 0.04$ & $-0.14 \pm 0.01$  \\
  		\hline
	\end{tabular}
\end{table*}

\begin{table*}
	\centering
	\caption{Same as Table~\ref{tab:estimators_all} but the best-performing estimators based on 
	global properties + resolved optical IFU properties but no resolved 12 \micron.  Table~\ref{tab:estimators_no_12um_aco3p2} shows the best-fit results assuming $\alpha_\mathrm{CO} = 3.2$.}
	\label{tab:estimators_no_12um}
	\begin{tabular}{lllll|ccc|cc} %
		\hline
		& & & & & \multicolumn{3}{c|}{$\theta_i$ for pixel properties} & \multicolumn{2}{c|}{$\theta_i$ for global properties} \\
		\cmidrule(lr){6-8}  \cmidrule(lr){9-10} 
		RMS error & $n_\mathrm{gal}$ & $n_\mathrm{pix}$ & $\sigma_\mathrm{int}$ & Zero-point ($\theta_0$) &$\log \Sigma_*$ & $(12+\log \mathrm{O/H})$ & $\log \Sigma_\mathrm{SFR}$ & $\log \Sigma_\mathrm{NUV}$ & $b/a_\mathrm{disk}$ \\
		\hline
  $0.20$ &  $58$ &  $1126$ &  $0.19$ & $2.00 \pm 0.01$ &  -- &  -- & $0.50 \pm 0.01$ &  -- &  --  \\
  $0.20$ &  $42$ &  $942$ &  $0.18$ & $1.86 \pm 0.01$ &  -- &  -- & $0.50 \pm 0.01$ &  -- & $0.22 \pm 0.01$  \\
  $0.17$ &  $27$ &  $552$ &  $0.18$ & $1.3 \pm 0.1$ & $0.18 \pm 0.01$ &  -- & $0.35 \pm 0.01$ & $0.01 \pm 0.01$ &  --  \\
  $0.17$ &  $27$ &  $552$ &  $0.18$ & $8.0 \pm 0.6$ & $0.23 \pm 0.01$ & $-0.81 \pm 0.07$ & $0.32 \pm 0.01$ & $0.02 \pm 0.01$ &  --  \\
  		\hline
	\end{tabular}
\end{table*}

\subsection{Dependence of the 12 \micron-H$_2$ relationship on physical scale}

\begin{figure*}
	\includegraphics[width=2\columnwidth]{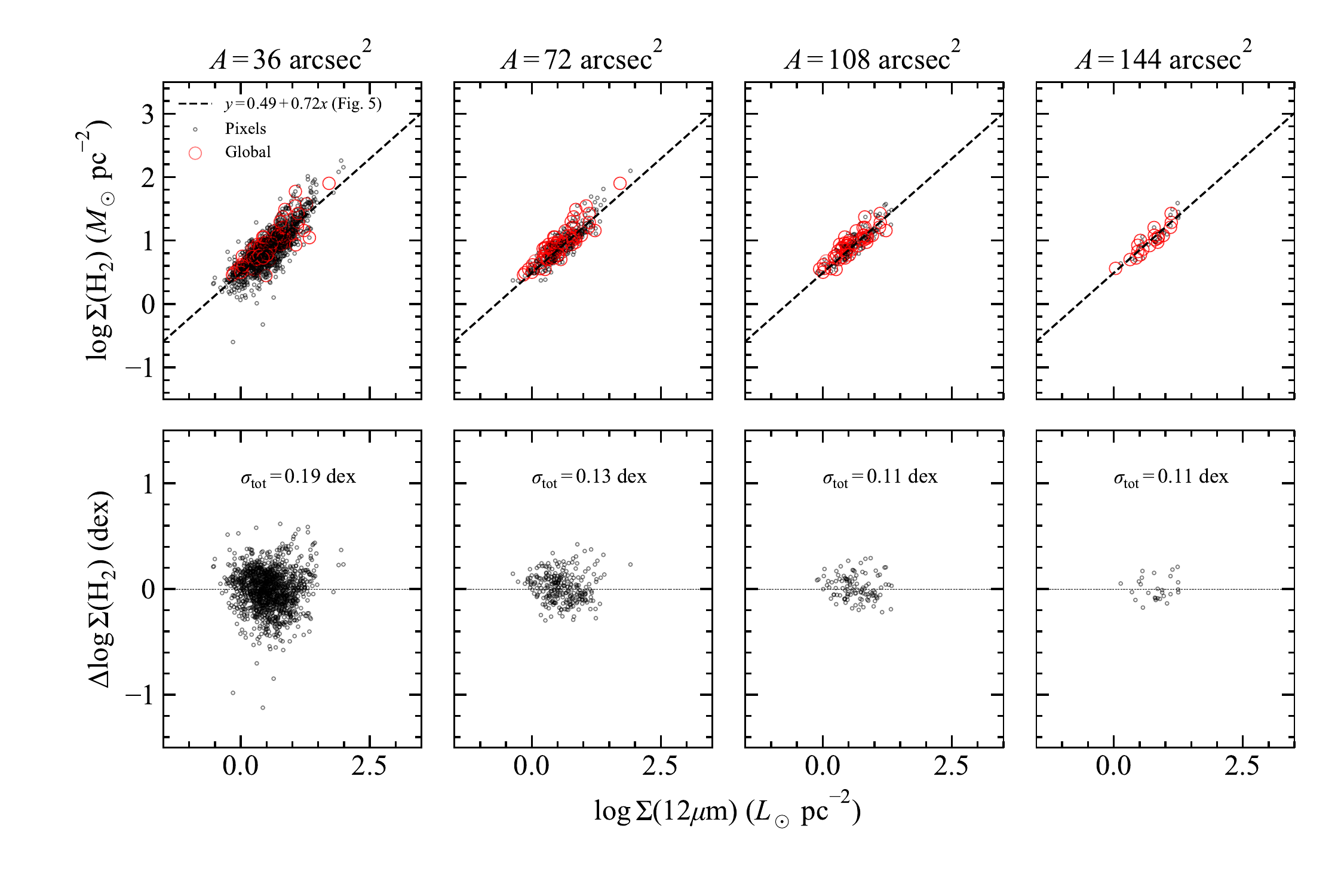}
    \caption{Variation of the scatter in the $\Sigma(\mathrm{H_2})$-$\Sigma(\mathrm{12\>\mu m})$ relationship with the area over which surface densities are calculated. \textit{Top:} black points are surface densities computed over area $A$ indicated at the top (36 arcsec$^2$ is one 6 arcsec pixel). Red circles are the sum of all pixels for each galaxy in the sample, and are the same in all panels in which that galaxy appears. The H$_2$ surface densities are computed with a metallicity-dependent $\alpha_\mathrm{CO}$. For each galaxy, all contiguous CO-detected, star-forming pixels with area $A$ were used. Each pixel was used exactly once in each panel from left to right. The number of galaxies decreases from left to right because some galaxies do not have any contiguous pixels which form the specified area. The fit to individual pixels is the same in all panels. \textit{Bottom:} residuals in 12 \micron\ surface density, relative to the resolved pixel fit (black line) from the bottom right panel of Figure~\ref{fig:w3_vs_co}. The total scatter $\sigma_\mathrm{tot}$ about the resolved fit decreases as the surface area approaches the total galaxy area, suggesting that the global correlation (red circles) emerges from the resolved correlation (black circles).}
    \label{fig:w3_vs_co_n}
\end{figure*}

To establish whether the correlation between global surface densities (12 \micron\ vs H$_2$) 
arises from a local correlation between pixel-based surface densities, 
we computed residuals of the individual pixel measurements from the resolved pixel fit
(right panel of Figure~\ref{fig:w3_vs_co}) with varying surface areas (Figure~\ref{fig:w3_vs_co_n}). 
For each galaxy, contiguous regions of 1, 4, 7 or 9 pixels were used to compute surface densities (the four columns of Figure~\ref{fig:w3_vs_co_n}).
The contiguous pixels were required to be CO-detected and star-forming, 
as a metallicity-dependent $\alpha_\mathrm{CO}$ was used. Each pixel was used in exactly one
surface density calculation for each resolution, so all of the black circles are independent.
We found that the scatter diminished as
the pixel size approached the whole galaxy size. 
The total scatter about the individual pixel fit declines as pixel area increases, indicating that 
the global correlation emerges from the local one.

\subsection{Testing the estimators for biases} \label{sec:bias}

To determine whether the best-fit relations are biased with respect to
any global or resolved properties (Table~\ref{tab:fitprops}), we 
performed the following tests for the best-performing H$_2$ estimators with 1, 2, and 3 parameters from Table~\ref{tab:estimators_all}.

For resolved properties, we plotted the residual in predicted vs. true $\log\Sigma(\mathrm{H_2})$ for each pixel versus 
resolved properties. We computed the Pearson-$r$ between the residuals and the resolved quantities.
No significant correlations were found for any of the resolved properties.
This indicates that the performance of the estimators is not biased with respect to resolved properties.

For global properties, we plotted the RMS error (Equation~\ref{eq:rmse}) for each galaxy versus 
global properties for that galaxy. We computed the Pearson-$r$ between the RMS error and  global quantities.
No significant correlations were found for any of the global properties. This indicates that the performance of the estimators is not biased with respect to global properties.

\section{Discussion}

Our findings show that significant power-law correlations between 
12 \micron\ and CO surface densities at kiloparsec scales 
are responsible for the observed correlation between global (galaxy-wide)
measurements \citep{jiang2015, gao2019}.
The median correlation coefficient between $\log \Sigma(\mathrm{12\>\micron})$ and 
$\log \Sigma(\mathrm{H_2})$ is $\simeq0.86$ (per galaxy).
Linear fits for each galaxy yield a range of intercepts 
spanning $\simeq1$ dex 
($-0.31$ to $0.87$, median 0.41),
and a range in slopes 
(0.20 to 2.03, median 1.13).
The 12 \micron\ and CO luminosities computed over the CO-detected area of each galaxy 
in the sample are well-fit by a single power law, with a larger slope
and smaller y-intercept than the fit to all individual-pixel 
luminosities in the sample. 
Linear regression on all possible combinations of resolved properties and global properties
(Table~\ref{tab:fitprops}) yielded several equations which can be used to 
estimate $\Sigma (\mathrm{H_2})$ (assuming a metallicity-dependent $\alpha_\mathrm{CO}$) in individual pixels.
A catalog of all resolved and global properties for each pixel in the analysis 
is provided in machine-readable format (Table~\ref{tab:catalog}).
The estimators were ranked according to the average accuracy with which
they can predict $\Sigma (\mathrm{H_2})$ in a given pixel (RMS error, Eq.~\ref{eq:rmse}).
The best-performing estimators (Tables~\ref{tab:estimators_all},~\ref{tab:estimators_no_ifu},~\ref{tab:estimators_no_12um}) with 1-4 independent variables 
are provided, and there is only marginal improvement in prediction error beyond 3 independent variables.
Out of all possible parameter combinations considered, 
the best-performing estimators include resolved $\Sigma (\mathrm{12\>\mu m})$, 
indicating that 12 \micron\ emission is likely physically linked to H$_2$ at resolved scales.

\begin{landscape}
\begin{table}
\caption{Selected rows and columns of the catalog of resolved measurements for each pixel considered in the analysis. 
A full version with more columns and rows is available in machine-readable format. 
A Python script is provided which shows how to reconstruct two-dimensional
images of all quantities in the catalog for each galaxy.
The luminosities corresponding to the surface densities in
columns 9-12 are provided in the full catalog.}
\label{tab:catalog}
\begin{tabular}{rlcccccccccc}
\hline
Pixel ID &        Galaxy &    BPT & $12+\log \mathrm{O/H_{pix}}$ & $\alpha_\mathrm{CO}$ & $\log \Sigma_{*,\mathrm{pix}}$ & $\log \Sigma_\mathrm{SFR,pix}$ & $A_{V,\mathrm{pix}}$ & $\log \Sigma_\mathrm{H_2}$ (Simple) & $\log \Sigma_\mathrm{H_2}$ (Sun) & $\log \Sigma_\mathrm{H_2}$ (Sun, $\alpha_\mathrm{CO}(Z)$) & $\log \Sigma(\mathrm{12\>\mu m})$ \\
(1) & (2) & (3) & (4) & (5) & (6) & (7) & (8) & (9) & (10) & (11) & (12) \\
\hline
    1464 &       NGC5980 &  Comp. &                           -- &                   -- &                         $2.49$ &                             -- &                   -- &                       $1.27\pm0.06$ &                    $1.24\pm0.05$ &                                                 -- &                   $1.09\pm0.02$ \\
    1465 &       NGC5980 &  Comp. &                           -- &                   -- &                         $3.13$ &                             -- &                   -- &                       $1.68\pm0.05$ &                    $1.70\pm0.03$ &                                                 -- &                   $1.21\pm0.02$ \\
    1466 &       NGC5980 &     SF &                         8.83 &                 2.44 &                         $2.48$ &                 $-1.61\pm0.03$ &               $1.08$ &                       $1.06\pm0.11$ &                    $1.13\pm0.06$ &                                      $1.01\pm0.06$ &                   $1.03\pm0.02$ \\
    1467 &       NGC5980 &     SF &                         8.84 &                 2.40 &                         $1.60$ &                 $-2.10\pm0.02$ &               $1.02$ &                            $< 1.09$ &                    $0.65\pm0.07$ &                                      $0.52\pm0.07$ &                   $0.65\pm0.02$ \\
    1468 &       NGC5980 &  Comp. &                           -- &                   -- &                         $0.73$ &                             -- &                   -- &                            $< 1.12$ &                               -- &                                                 -- &                   $0.17\pm0.02$ \\
    1469 &       NGC5980 &     SF &                         8.84 &                 2.39 &                        $-1.04$ &                 $-3.77\pm0.03$ &              $-3.02$ &                            $< 1.24$ &                               -- &                                                 -- &                  $-0.29\pm0.04$ \\
   24622 &       NGC4047 &     SF &                         8.80 &                 2.70 &                         $2.30$ &                 $-1.51\pm0.03$ &               $1.26$ &                       $1.58\pm0.09$ &                    $1.61\pm0.04$ &                                      $1.54\pm0.04$ &                   $1.17\pm0.02$ \\
   24623 &       NGC4047 &     SF &                         8.76 &                 2.97 &                         $2.65$ &                 $-1.30\pm0.02$ &               $1.21$ &                       $1.78\pm0.05$ &                    $1.87\pm0.03$ &                                      $1.83\pm0.03$ &                   $1.35\pm0.02$ \\
   24624 &       NGC4047 &     SF &                         8.71 &                 3.45 &                         $2.72$ &                 $-1.20\pm0.02$ &               $1.08$ &                       $1.88\pm0.05$ &                    $1.90\pm0.03$ &                                      $1.93\pm0.03$ &                   $1.41\pm0.02$ \\
   24625 &       NGC4047 &     SF &                         8.76 &                 3.02 &                         $2.46$ &                 $-1.40\pm0.02$ &               $1.11$ &                       $1.81\pm0.07$ &                    $1.81\pm0.03$ &                                      $1.78\pm0.03$ &                   $1.36\pm0.02$ \\
   24626 &       NGC4047 &     SF &                         8.83 &                 2.47 &                         $2.13$ &                 $-1.65\pm0.02$ &               $1.09$ &                       $1.47\pm0.13$ &                    $1.55\pm0.04$ &                                      $1.43\pm0.04$ &                   $1.20\pm0.02$ \\
   24627 &       NGC4047 &     SF &                         8.85 &                 2.32 &                         $2.04$ &                 $-1.82\pm0.03$ &               $1.28$ &                            $< 1.63$ &                    $1.20\pm0.08$ &                                      $1.06\pm0.08$ &                   $0.96\pm0.02$ \\
   24628 &       NGC4047 &  Comp. &                           -- &                   -- &                         $1.41$ &                             -- &                   -- &                            $< 1.63$ &                               -- &                                                 -- &                   $0.63\pm0.02$ \\
   24629 &       NGC4047 &     SF &                         8.80 &                 2.64 &                         $1.08$ &                 $-3.01\pm0.09$ &               $0.50$ &                            $< 1.65$ &                               -- &                                                 -- &                   $0.26\pm0.03$ \\
   24630 &       NGC4047 &     -- &                           -- &                   -- &                             -- &                             -- &                   -- &                            $< 1.72$ &                               -- &                                                 -- &                  $-0.06\pm0.05$ \\
   24631 &       NGC4047 &     -- &                           -- &                   -- &                             -- &                             -- &                   -- &                            $< 1.86$ &                               -- &                                                 -- &                  $-0.30\pm0.07$ \\
   24632 &       NGC4047 &     -- &                           -- &                   -- &                             -- &                             -- &                   -- &                            $< 1.71$ &                               -- &                                                 -- &                  $-0.52\pm0.12$ \\
   24633 &       NGC4047 &     -- &                           -- &                   -- &                             -- &                             -- &                   -- &                            $< 1.70$ &                               -- &                                                 -- &                  $-0.31\pm0.08$ \\
   24634 &       NGC4047 &     -- &                           -- &                   -- &                             -- &                             -- &                   -- &                            $< 1.79$ &                               -- &                                                 -- &                  $-0.02\pm0.04$ \\
   24635 &       NGC4047 &     SF &                         8.79 &                 2.73 &                         $1.21$ &                 $-2.34\pm0.05$ &               $1.12$ &                            $< 1.67$ &                               -- &                                                 -- &                   $0.26\pm0.03$ \\
   24636 &       NGC4047 &     SF &                         8.84 &                 2.36 &                         $1.65$ &                 $-2.45\pm0.04$ &               $0.86$ &                            $< 1.58$ &                               -- &                                                 -- &                   $0.55\pm0.02$ \\
   24637 &       NGC4047 &     SF &                         8.88 &                 2.13 &                         $1.90$ &                 $-2.34\pm0.04$ &               $0.68$ &                            $< 1.58$ &                    $1.16\pm0.06$ &                                      $0.98\pm0.06$ &                   $0.90\pm0.02$ \\
\hline
\multicolumn{12}{l}{(3) BPT classification (Section~\ref{sec:califa}): starforming (``SF''), composite (``Comp.''), low-ionization emission region (``LIER''), or Seyfert (``Sy'').}\\
\multicolumn{12}{l}{(5) Metallicity-dependent $\alpha_\mathrm{CO}$ (Eq.~\ref{eq:alphaco_met}) in units of $\mathrm{M_\odot} \mathrm{(K\> km \> s^{-1} \> pc^2)^{-1}}$.}\\
\multicolumn{12}{l}{(6) Resolved stellar mass surface density (Sec.~\ref{sec:califa}) in units of $\mathrm{M_\odot} \> \mathrm{kpc}^{-2}$.}\\ 
\multicolumn{12}{l}{(7) Resolved SFR surface density (Equation~\ref{eq:sigma_sfr}) in units of $\mathrm{M_\odot} \> \mathrm{yr}^{-1} \> \mathrm{kpc}^{-2}$.}\\
\multicolumn{12}{l}{(8) Resolved extinction derived from the Balmer decrement, in units of mag (Equation~\ref{eq:a_ha}).}\\
\multicolumn{12}{l}{(9) H$_2$ surface density ($\mathrm{M_\odot}\>\mathrm{pc^{-2}}$) based on the ``Simple'' moment-0 map (Method 2, Section~\ref{sec:h2maps}). Method 1 is better at improving the SNR in each pixel, so detects more pixels than Method 2.}\\
\multicolumn{12}{l}{ A constant $\alpha_\mathrm{CO}$ is assumed, and 98\%\ confidence $3\sigma$ upper limits are shown for non-detections.}\\
\multicolumn{12}{l}{(10) H$_2$ surface density ($M_\odot\>\mathrm{pc^{-2}}$) from the moment-0 map made using the \citet{sun2018} method (Method 1), assuming a constant $\alpha_\mathrm{CO}=3.2$.}\\
\multicolumn{12}{l}{(11) Same as (10) but assuming a metallicity-dependent $\alpha_\mathrm{CO}$ and only using star-forming pixels.}\\
\multicolumn{12}{l}{(12) Resolved 12 \micron\ surface density in units of $\mathrm{L_\odot} \> \mathrm{pc}^{-2}$.}\\
\end{tabular}
\end{table}
\end{landscape}

\subsection{Comparisons to previous work}

Previous work on the 12 \micron-CO relationship has been primarily focused 
on the total 12 \micron\ luminosity and the total CO luminosity for each galaxy
\citep{jiang2015, gao2019}.
Our fit of the global CO luminosity versus 12 \micron\ luminosity over the CO-detected area
(Figure~\ref{fig:w3_vs_co}) yields 
a slope of $0.94\pm0.04$ and intercept of $0.46\pm0.38$.
Our slope agrees well with \citet{gao2019} who find $0.98\pm0.02$, but our intercept 
is significantly greater than their value of $-0.14 \pm 0.18$.
Our global CO luminosities are
consistent with those reported in \citetalias{bolatto2017}, which 
are believed to be accurate estimates of the true total CO luminosities (see Section 3.2 in \citetalias{bolatto2017}).
However, we find that our global 12 \micron\ luminosities (the sum over the CO-detected area) 
are systematically
lower than the true total 12 \micron\ luminosities as measured by the method in
\citet{gao2019}. The amount of discrepancy is consistent with the offset in intercept
found between this work and \citet{gao2019}.
This comparison indicates that 12 \micron\ emission tends to be more spatially extended than CO emission, 
so by restricting the area to the CO-emitting area, some 12 \micron\ emission is missed, leading to a smaller intercept.
The fact that this does not affect the slope indicates that the fraction of 12 \micron\ emission 
that is excluded by only considering the CO-detected area, 
is similar from galaxy to galaxy.

When estimating the total CO luminosity in a galaxy, we recommend 
cross-checking with the \citet{gao2019} estimators because they take the total 12 \micron\
luminosity as input, whereas our estimators require the 12 \micron\ 
luminosity \textit{over the CO-detected area}.
Since our total CO luminosities
agree with the total CO luminosities presented in \citetalias{bolatto2017},
it is unlikely that these interferometric measurements significantly underestimate
the true total CO luminosities. However, since a comparison of the EDGE 
total CO luminosities with single-dish measurements for the same sample 
has not been done, it is not impossible that there is some missing flux.

Our results can be compared to recent work using optical extinction 
as an estimator of H$_2$ surface density \citep{guver2009, barrera-ballesteros2016, concas2019, yesuf2019, barrera-ballesteros2020}. We show that 
resolved 12 \micron\ surface density is better than optical extinction at predicting H$_2$ surface density
by $\simeq0.1$ dex per pixel (Figure~\ref{fig:rmse}).
Additionally, a 12 \micron\ estimator does not suffer from 
a limited dynamic range like $A_V$ traced by the Balmer decrement,
which is invalid at large extinctions,
and where the SNR of the H$\alpha$ and H$\beta$ lines are low.
In the recent analysis of EDGE galaxies \citet{barrera-ballesteros2020} 
limit their analysis to $A_V < 3$ due to the SNR of the H$\beta$ line.
Additionally, the correlation between resolved $\Sigma(12\mathrm{\mu m})$ 
and $\Sigma(\mathrm{H_2})$ is stronger than that between 
$A_V$ and $\Sigma(\mathrm{H_2})$.

\subsection{Why is $\Sigma$(12 \micron) a better predictor of $\Sigma$(H$_2$) than $\Sigma_\mathrm{SFR}$?}

Over the same set of pixels (star forming and CO detected), the correlation 
between $\log\Sigma(\mathrm{12\>\mu m})$ and $\log\Sigma(\mathrm{H_2})$ per galaxy (left panel, Figure~\ref{fig:pearson_hist})
is better than the correlation between $\log\Sigma_\mathrm{SFR}$ and $\log\Sigma(\mathrm{H_2})$ (right panel, Figure~\ref{fig:pearson_hist}). This is also apparent
from our findings that estimators of $\Sigma($H$_2)$ based on $\Sigma(\mathrm{12\>\mu m})$ consistently perform better 
at predicting $\Sigma(\mathrm{H_2})$ than estimators with $\Sigma_\mathrm{SFR}$ instead of $\Sigma(\mathrm{12\>\mu m})$ (Section~\ref{sec:estimator}). 

Since we have restricted our analysis to star-forming pixels, 
the 12 \micron\ emission that we see 
is likely dominated by the 11.3 \micron\ PAH feature.
The underlying continuum emission can arise 
from warm, very small dust grains 
heated by AGN. This likely does not dominate the 12 \micron\ 
emission since most ($\sim 80$ per cent) of the \wise\ 12 \micron\ emission 
in star-forming galaxies is from stellar populations younger than 0.6 Gyr \citep{donoso2012}.
However, it is important to rule out any effects of obscured AGN.
PAH emission is known to be affected by the presence of an AGN 
\citep{diamond-stanic2010, shipley2013, jensen2017, alonso-herrero2020}, 
but there is conflicting evidence on the nature of this relationship.
For example, \citet{tommasin2010} find AGN-dominated and starburst-dominated galaxies have roughly the same 11.3 \micron\ PAH flux, while \citet{murata2014} and \citet{maragkoudakis2018} find suppressed PAH
emission in starburst galaxies relative to galaxies with AGN. In contrast, 
\citet{shi2009} and \citet{shipley2013} find suppressed PAH emission in AGN compared to non-AGN.
If there are any obscured AGN in our sample, they would not be identified as AGN
from the BPT method. However, since our pixels are 1 to 2 kpc in size,
the impact of an obscured AGN would be 
restricted to the central pixel of the galaxy.
To assess the potential impact of obscured AGN on our results, 
we redid all of our multiparameter fits with the central pixel of 
each galaxy masked if it was not already masked based on the BPT classification.
We found that the 12 \micron-H$_2$ correlation remains stronger than the SFR-H$_2$ correlation,
and that the fit parameters do not change significantly (they are consistent within
the quoted uncertainties).
Thus we are confident that AGN do not significantly impact our results.

These results have implications for the connection between 
emission that is traced by the 12 \micron\ band (mostly PAHs) and CO emission. 
Exactly how and where PAHs are formed is not currently understood 
\citep[for a recent review from the \textit{Spitzer} perspective see][]{li2020a}, but traditionally PAHs have been modelled to absorb FUV photons through the photoelectric effect and eject electrons into the ISM, which heats the gas \citep{bakes1994, tielens2008}. 
Since PAHs are excited by stellar UV photons, PAH emission has been 
considered as an SFR tracer \citep[e.g.][]{roussel2001, peeters2004, wu2005, shipley2016, cluver2017, xie2019, whitcomb2020}.
Although the PAH-SFR connection 
breaks down at sub-kpc scales
\citep{werner2004, bendo2020}, PAH emission is still used as an SFR tracer on
global scales for low-redshift galaxies \citep{kennicutt2009, shipley2016}.
\wise\ 12 \micron\ emission has also been examined as a SFR indicator; however
its relationship with SFR shows greater scatter than the \wise\ 22 \micron-SFR relationship \citep{jarrett2013, cluver2017, leroy2019}. Similar to the 8 \micron\ emission vs. SFR relation \citet{calzetti2007},
the complex relationship between thermal dust, PAH emission and star formation activity
adds scatter to the correlations between MIR emission and SFR \citep{jarrett2013}.

Many studies have also found that there is a tight link between PAHs and 
the contents of the interstellar medium: molecular gas traced by CO \citep{regan2006, pope2013, cortzen2019}, 
and cold ($T\sim 25$ K) dust, which traces the bulk of the ISM
\citep{haas2002, bendo2008, jones2015, bendo2020}.
Milky Way studies have found that PAH emission is enhanced surrounding and suppressed within \ion{H}{II} regions \citep[e.g.][]{churchwell2006, povich2007}.
In addition, the PdBI Arcsecond Whirlpool Survey \citep[PAWS;][]{schinnerer2013} 
of cold gas in M51 with cloud-scale resolution ($\sim 40$ pc) found that \textit{Spitzer} 8 \micron\ PAH emission and CO(1-0) emission
are highly correlated in position but not in flux, and that most of the PAH emission appears to be coming from only the surfaces of giant molecular clouds.
These results and others such as \citet{sandstrom2010} suggest that PAHs are either formed in molecular clouds
or destroyed in the diffuse ISM, and that 
the conditions of PAH formation and CO formation are likely similar.
The suppression of PAH emission in \ion{H}{II} regions 
may be due to decreased dust shielding, analogous to how CO emission
is reduced in low-metallicity regions, or to 
changes in how PAHs are formed and/or destroyed \citep{sandstrom2013, li2020a}.
It is plausible that 
our findings support a picture in which PAHs form in molecular clouds or are destroyed in the diffuse ISM; however due to the difference
in physical resolution, and the contribution of continuum emission and multiple PAH features
to the 12 \micron\ emission, a study focused specifically on 11.3 \micron\ PAH instead of
\wise\ 12 \micron\ would be required.
Overall it seems likely that the strength of the 12 \micron-H$_2$ 
correlation in star-forming regions is due to the combination 
of the Kennicutt-Schmidt relation and a 
direct link between the 11.3 \micron\ PAH feature and molecular gas as traced by CO.

\section{Conclusions}

We find that \wise\ 12 \micron\ emission and CO(1-0)
emission from EDGE are highly correlated at $\sim$ kpc scales in star-forming regions of nearby galaxies after matching 
the resolution of the two data sets. Using multi-variable linear regression 
we compute linear combinations of resolved and global galaxy properties
that robustly predict H$_2$ surface densities. We find that 12 \micron\
is the best predictor of H$_2$, and is notably better than $\Sigma_\mathrm{SFR}$ derived from
resolved H$\alpha$ emission. Our results are statistically robust, and are not significantly affected by
the possible presence of any obscured AGN or by assumptions about the CO-to-H$_2$ conversion factor.
We interpret these findings as further evidence that  
11.3 \micron\ PAH emission is more spatially correlated with H$_2$ than with \ion{H}{II} regions.
Although the details of the life cycle and excitation of PAH molecules are 
not fully understood,
we believe that the strong correlation between 12 \micron\ and CO emission 
is likely due to the 
fact that PAH emission is both a SFR tracer and a cold ISM tracer. 
Additionally, if PAHs are indeed formed within molecular clouds and in similar conditions 
to CO as previous work suggests,
we suspect that the \wise\ 12 \micron-CO correlation will persist at molecular-cloud scale
resolution.

We present resolved $\Sigma_\mathrm{H_2}$ estimators which
can be used for two key applications:
\begin{enumerate}
\item generating large samples of estimated resolved $\Sigma(\mathrm{H_2})$ in the nearby Universe e.g. to study the resolved Kennicutt-Schmidt law, and
\item predicting $\Sigma(\mathrm{H_2})$ and integration times for 
telescope observing proposals (e.g. ALMA). 
\end{enumerate}
Although the CO-detected pixels in our sample only extend down to
$\Sigma(\mathrm{H_2})\sim 1 \> \mathrm{M_\odot} \> \mathrm{pc^{-2}}$, our predictions for $\Sigma(\mathrm{H_2})$ below this are consistent with the upper limits in our data. However, we advise caution when applying the estimator to 12 \micron\ surface densities below about $1 \> \mathrm{L_\odot} \> \mathrm{pc^{-2}}$.
Since \wise\ was an
all-sky survey, in principle these estimators could be applied over the entire sky.
In the future, using the MIR data with higher resolution and better sensitivity from the \textit{James Webb Space Telescope} instead of \wise\ 12 \micron, and ALMA CO data instead of CARMA CO data, one could produce 
an H$_2$ surface density estimator which reaches even lower gas surface densities.

\section*{Acknowledgements}

We thank the anonymous referee for his/her suggestions that have improved the manuscript.
CL acknowledges the support by the National Key R\&D Program of China
(grant No. 2018YFA0404502, 2018YFA0404503), and the National 
Science Foundation of China (grant Nos. 11821303, 11973030, 11673015, 
11733004, 11761131004, 11761141012). YG acknowledges funding from National Key Basic Research and Development Program of China (grant No. 2017YFA0402704). LCP and CDW acknowledge support from the Natural Science and Engineering 
Research Council of Canada and CDW acknowledges support from the Canada Research Chairs program.

This publication makes use of data products from the 
\textit{Wide-field Infrared Survey Explorer}, 
which is a joint project of the University of California, Los Angeles, 
and the Jet Propulsion Laboratory/California Institute of Technology, 
funded by the National Aeronautics and Space Administration.
This research has made use of the NASA/IPAC Infrared Science Archive, which is funded by the National Aeronautics and Space Administration and operated by the California Institute of Technology.
This study uses data provided by the Calar Alto Legacy Integral Field Area (CALIFA) survey (\url{http://califa.caha.es/}). Based on observations collected at the Centro Astronomico Hispano Aleman (CAHA) at Calar Alto, operated jointly by the Max-Planck-Institut fur Astronomie and the Instituto de Astrofisica de Andalucia (CSIC).
We acknowledge the usage of the HyperLEDA database (\url{http://leda.univ-lyon1.fr}).
This research was enabled in part by support provided by WestGrid (\url{https://www.westgrid.ca}) and Compute Canada (\url{https://www.computecanada.ca}).

\section*{Data Availability}

The data underlying this article are available in the article and in its online supplementary material.

\bibliographystyle{mnras}
\bibliography{wise_co_paper_v2} %

\appendix

\section{Derivation of \wise\ W3 uncertainty} \label{appendix:w3unc}

The \textit{total} uncertainty in each 6 arcsec pixel is 
the instrumental uncertainty added in quadrature with the 
zero-point uncertainty
\begin{equation}
\sigma_\mathrm{12\>\mu m, \> tot} = \sqrt{\sigma_\mathrm{inst.,\> final}^2 + \sigma_\mathrm{ZP}^2}.
\end{equation}
The instrumental uncertainty in each pixel was measured by taking
the uncertainty maps from the \wise\ archive, adding the native pixels in 
quadrature into 6 arcsec pixels, taking the square root, and multiplying the
resulting map by the unit conversion factor in Equation~\ref{eq:w3units}.
The instrumental noise variance in each larger pixel is
\begin{equation}
\sigma_\mathrm{inst.,\> final}^2 = 5 \sum_\mathrm{subpixels} \sigma_\mathrm{inst., \> natv.}^2,
\end{equation}
where the factor of 5 correction was estimated from Figure 3 of 
\url{http://wise2.ipac.caltech.edu/docs/release/allsky/expsup/sec2_3f.html} (since 
our 6 arcsec pixels are effectively apertures with radius of $3/1.175 = 2.5$ pixels),
and $\sigma_\mathrm{inst., \> natv.}$ is the instrumental uncertainty at the native pixel scale.

There is a 4.5 per cent uncertainty in the W3 zero-point
magnitude \citep[Figure 9 of][]{jarrett2011}, such that
\begin{equation}
\sigma_\mathrm{MAG} = \frac{2.5}{\ln 10} \frac{\sigma_F}{F} = 0.045,
\end{equation}
or $\sigma_F = 0.0414 F$.
The zero-point uncertainty is given by
\begin{equation}
\sigma_\mathrm{ZP} = 0.0414 \sum_\mathrm{subpixels} F_\mathrm{natv.},
\end{equation}
where $F_\mathrm{natv.}$ is the 
flux at the native pixel scale.

\section{Derivation of CO uncertainty} \label{appendix:counc}

A noise map $N(x,y)$ (in $\mathrm{Jy\>beam^{-1}\>km\>s^{-1}}$) is calculated by adding 
a 10 per cent calibration uncertainty in quadrature with the instrumental uncertainty
\begin{equation}
\frac{N(x,y)}{\mathrm{Jy \>beam^{-1} \> km \>s^{-1}}} = \left\{ \left[0.1 M_0(x,y)\right]^2 + \sigma(x,y)^2 \frac{N_\mathrm{pix,beam}}{f_\mathrm{bin}}\right\}^{1/2},
\end{equation}
where $M_0(x,y)$ is the moment-0 map ($\mathrm{Jy\>beam^{-1}\>km\>s^{-1}}$) with 6 arcsec pixels, 
the factor of 0.1 is a 10 per cent calibration uncertainty,
$N_\mathrm{pix,beam}$ is the number of pixels per beam in the raw image 
(prior to any rebinning),
$f_\mathrm{bin}$ is the binning factor (the number of original pixels in the \textit{rebinned} pixels, e.g. since we went from $1\arcsec\times 1\arcsec$ to $6\arcsec \times 6\arcsec$ pixels,  $f_\mathrm{bin}=36$),
and
\begin{equation}
\frac{\sigma(x,y)}{\mathrm{Jy\>beam^{-1}\>km\>s^{-1}}} = \left(\frac{\Delta v_\mathrm{chan}}{\mathrm{km\>s^{-1}}}\right)\sqrt{N_\mathrm{chan}(x,y)} \left(\frac{\sigma_\mathrm{chan}}{\mathrm{Jy\>beam^{-1}}}\right),
\end{equation}
where $\Delta v_\mathrm{chan} = 20$ km s$^{-1}$ is the velocity width of the channels in the cube,
$N_\mathrm{chan}(x,y)$ is the number of channels used to calculate the moment-0 map 
(which varies with position),
and $\sigma_\mathrm{chan}$ is the RMS per channel.
When calculating upper limits, $N_\mathrm{chan}(x,y)=34$ for all pixels.
In a CO cube, $\sigma_\mathrm{chan}$ is calculated by measuring the RMS 
of all pixels within a 7 arcsec radius circular aperture in the 
center of the field in the first 
3-5 channels, and again in the last 3-5 channels. $\sigma_\mathrm{chan}$ is 
taken to be the average of these two RMSes.
Finally, we convert the noise maps into units of luminosity using Equation~\ref{eq:lco1}.

\section{Definition of the scatter about a fit}\label{appendix:scatter}

The total scatter about a fit $\sigma_\mathrm{tot}$ is
\begin{equation}\label{eq:sig_tot}
\sigma_\mathrm{tot} = \sqrt{\frac{1}{N-m}\sum_i(y_i-\hat{y_i})^2},
\end{equation}
where $N$ is the number of data points, $m$ is the number of fit parameters,
$y_i$ is $i$'th independent variable, and $\hat{y_i}$ is the estimate of $y_i$ 
from the fit. $\sigma_\mathrm{tot}$ can be directly computed from the fit.
The total scatter 
can also be written as the sum in quadrature of random scatter due to measurement 
uncertainties, and the remaining ``intrinsic'' scatter $\sigma_\mathrm{int}$
\begin{equation}
\sigma_\mathrm{tot} = \sqrt{\frac{1}{N}\sum_i\sigma_i^2 + \sigma_\mathrm{int}^2},
\end{equation}
where $\sigma_i$ is the measurement error on $y_i$.
The intrinsic scatter can be computed using 
\begin{equation}\label{eq:sig_int}
\sigma_\mathrm{int}  = \sqrt{\sigma_\mathrm{tot}^2 - \frac{1}{N}\sum_i\sigma_i^2}.
\end{equation}

\section{The 12 \micron-CO relationship assuming a constant $\alpha_\mathrm{CO}$}\label{appendix:w3_co_v2}

\begin{figure*}
	\includegraphics[width=\columnwidth]{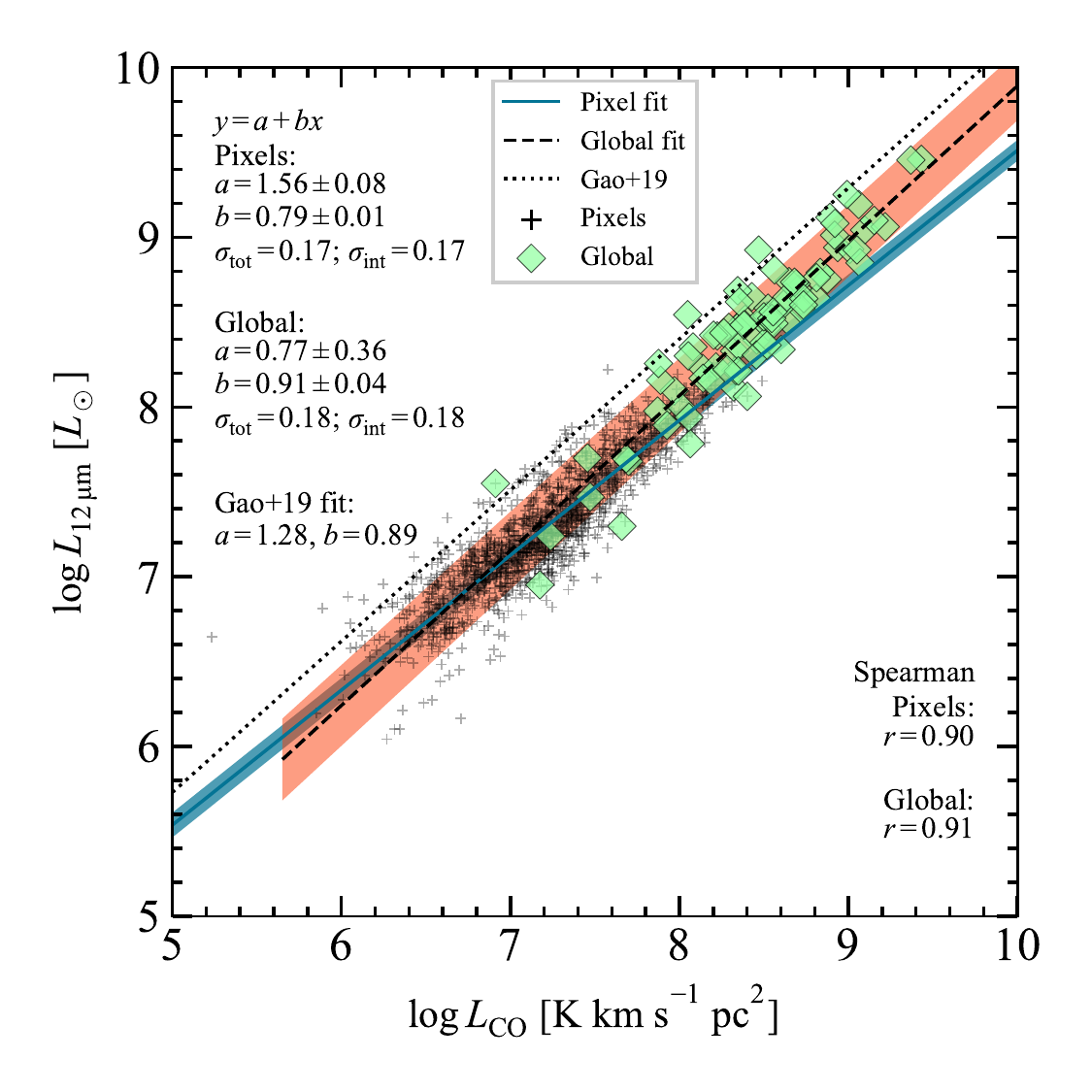}\includegraphics[width=\columnwidth]{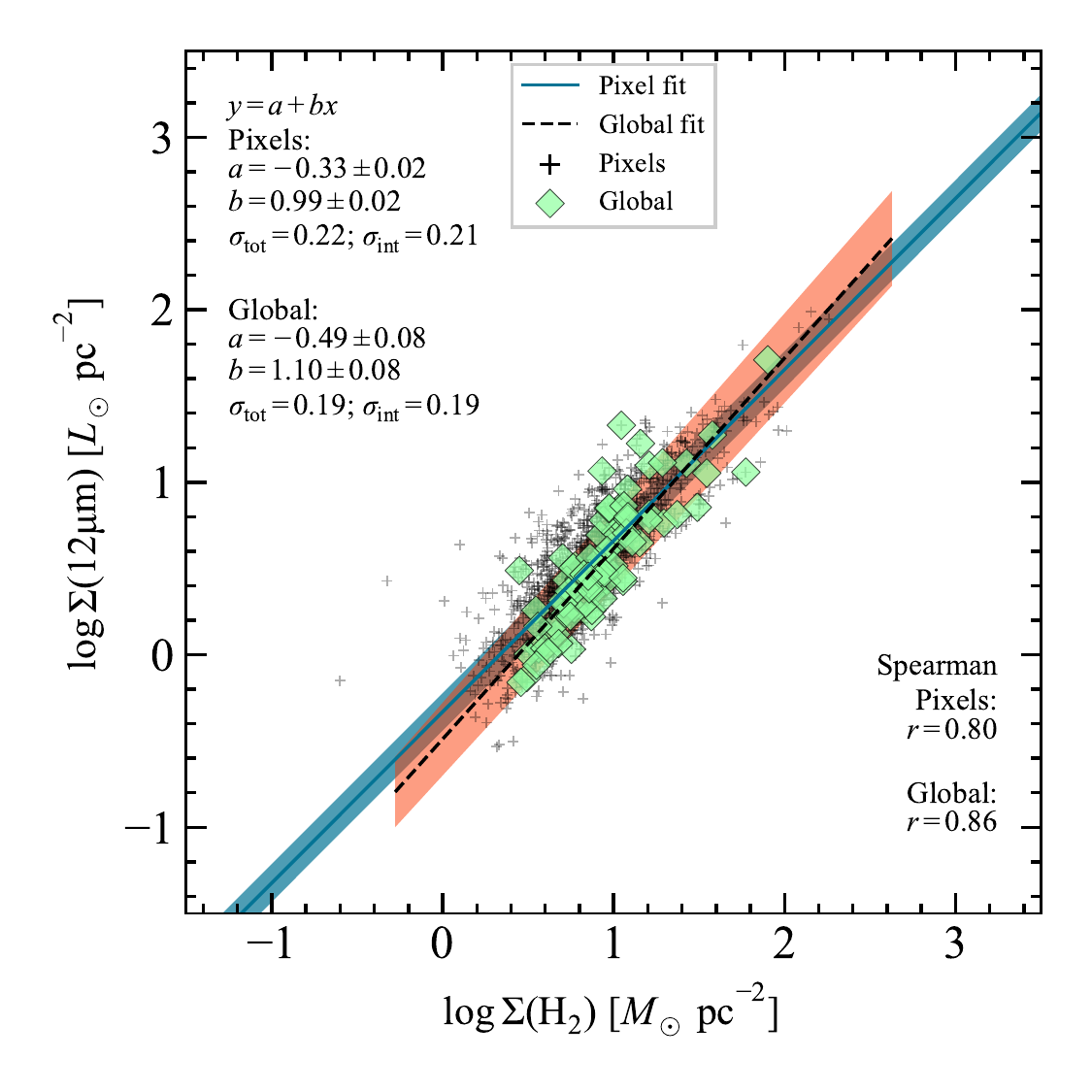}
    \caption{
    Same as Figure~\ref{fig:w3_vs_co} but with the x and y axes interchanged.
}
    \label{fig:w3_vs_co_v3}
\end{figure*}

Figure~\ref{fig:w3_vs_co_v3} shows the 12 \micron\ vs. CO relationship 
in terms of luminosities (left) and surface densities (right), as in Figure~\ref{fig:w3_vs_co}
except with the x and y axes interchanged, and the fits redone.

\begin{figure*}
	\includegraphics[width=\columnwidth]{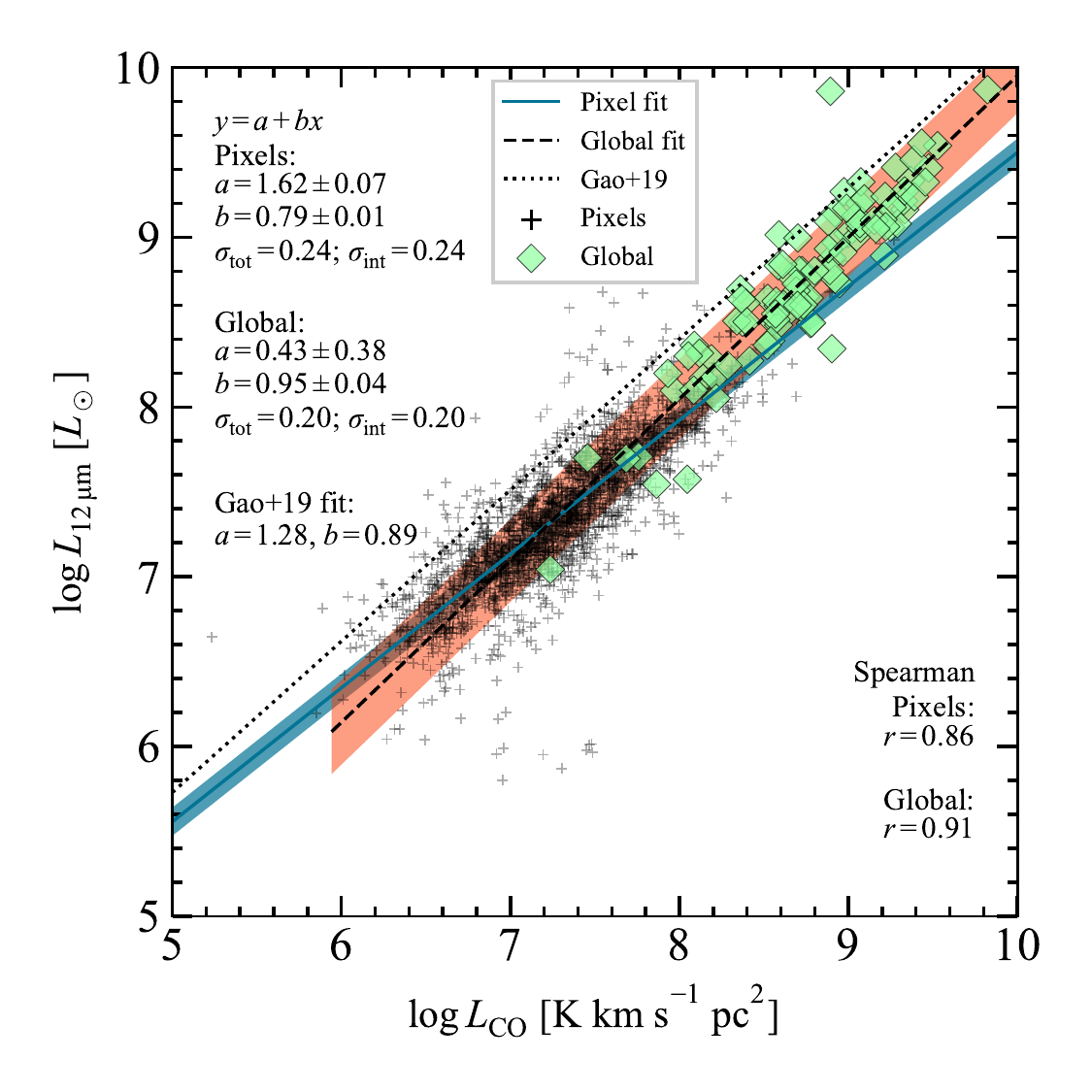}\includegraphics[width=\columnwidth]{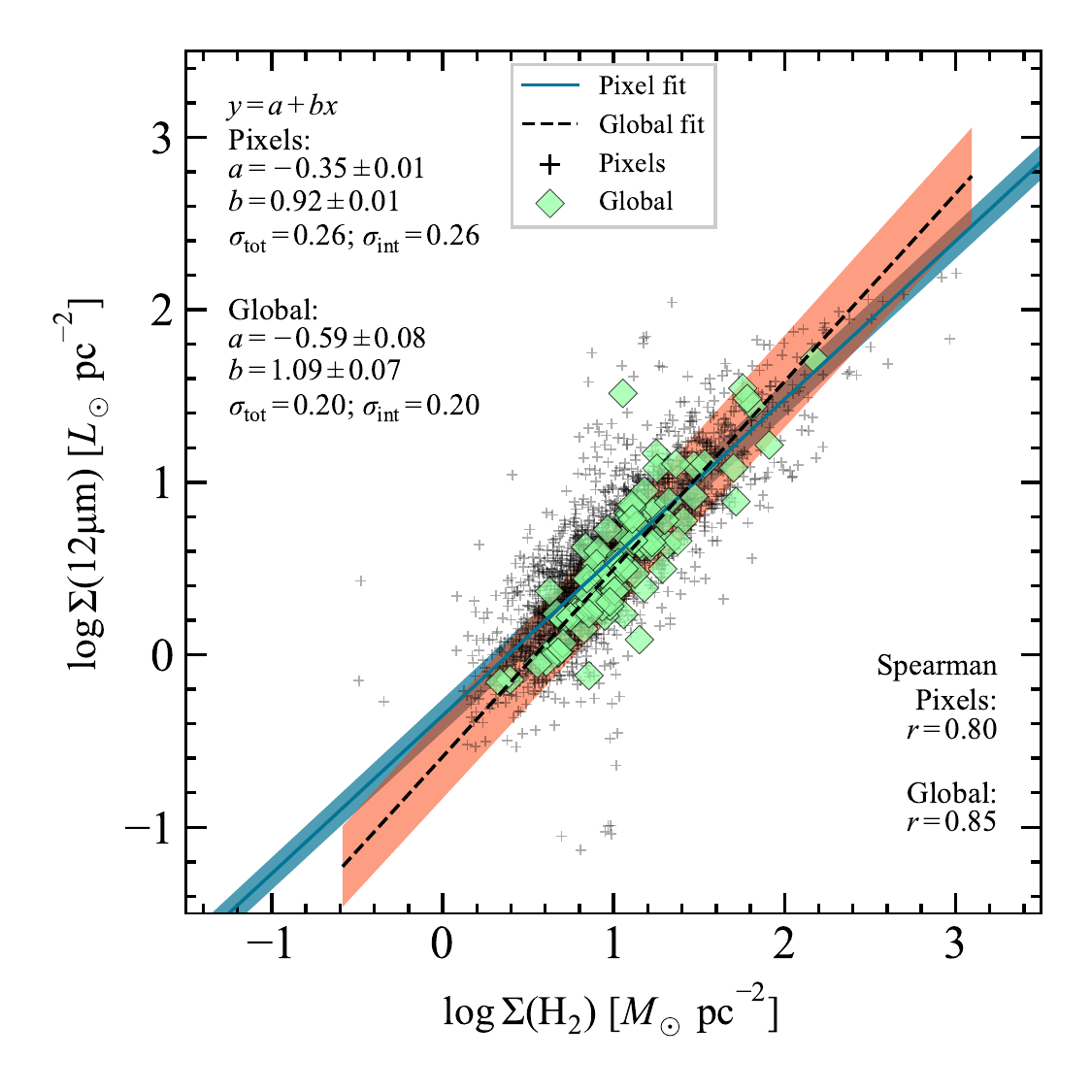}
	\includegraphics[width=\columnwidth]{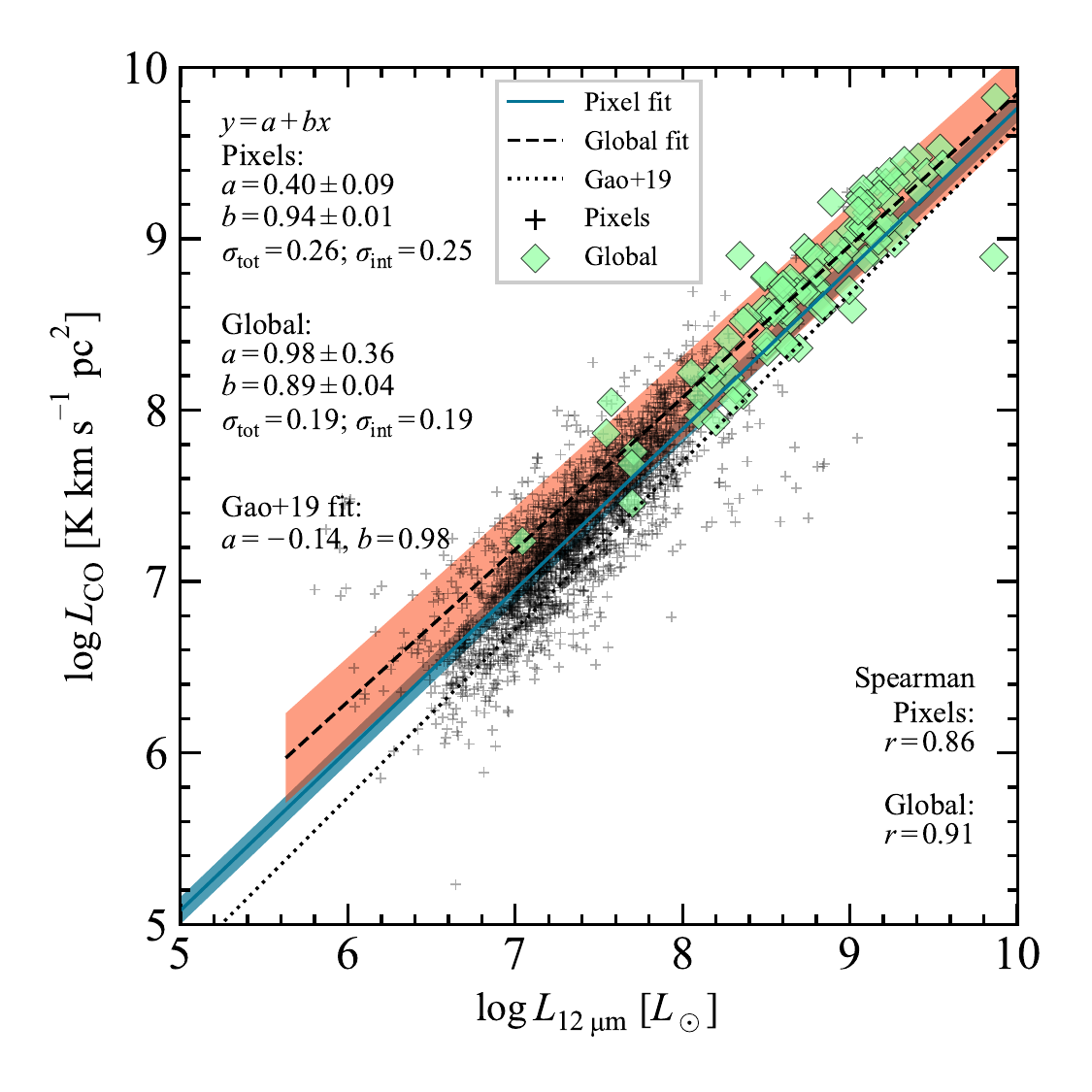}\includegraphics[width=\columnwidth]{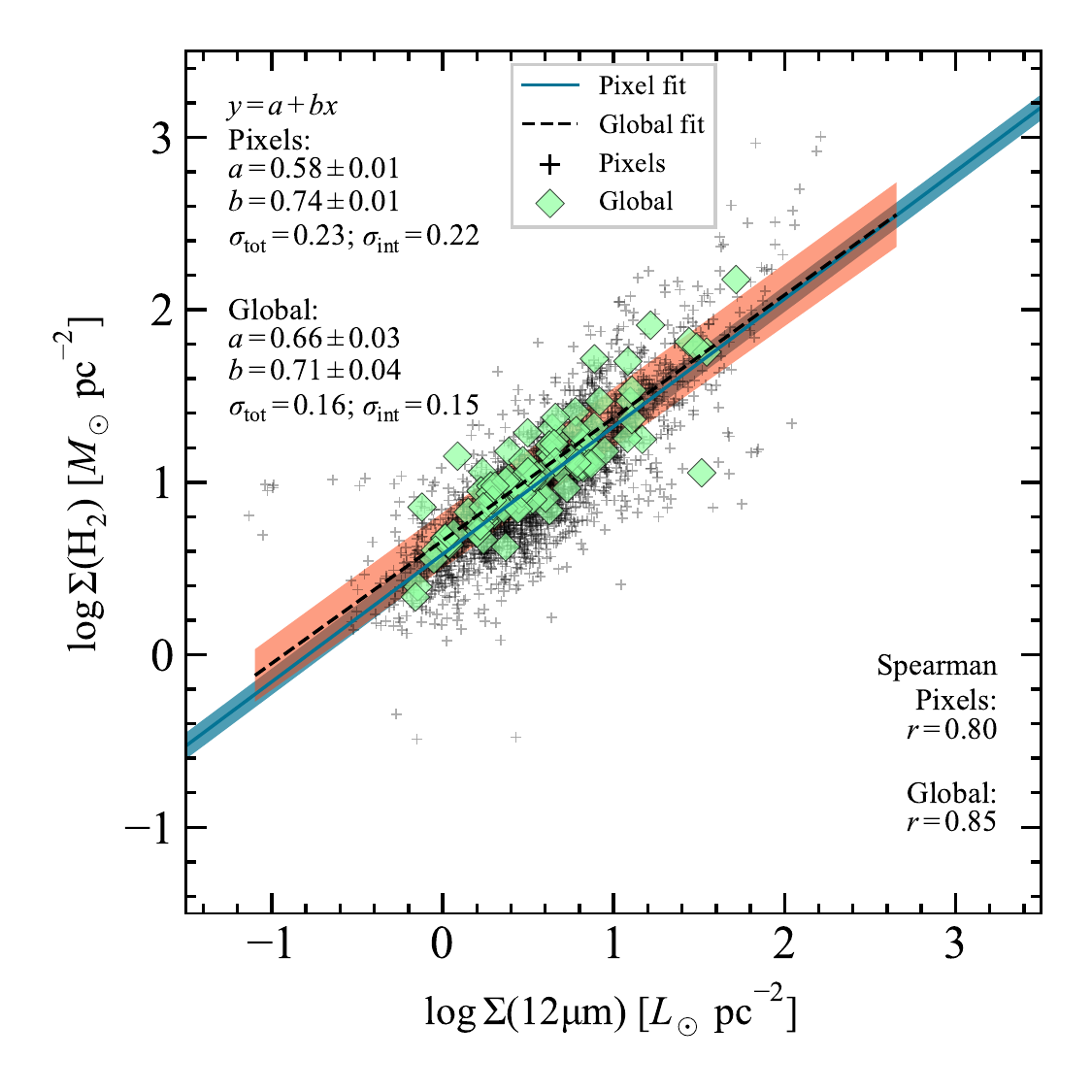}
    \caption{ Same as Figure~\ref{fig:w3_vs_co} except H$_2$ surface densities were calculated using $\alpha_\mathrm{CO}=3.2 \> \mathrm{M_\odot} \mathrm{(K\> km \> s^{-1} \> pc^2)^{-1}}$, and non-starforming pixels were included. }
    \label{fig:w3_vs_co_v2}
\end{figure*}

For completeness, Figure~\ref{fig:w3_vs_co_v2} shows the relationships and fits 
as Figure~\ref{fig:w3_vs_co} except assuming a constant CO-to-H$_2$ conversion factor
$\alpha_\mathrm{CO}=3.2 \> \mathrm{M_\odot} \mathrm{(K\> km \> s^{-1} \> pc^2)^{-1}}$,
and including all CO-detected pixels (not just star-forming). The changes from 
Figure~\ref{fig:w3_vs_co} are slight overall, and are the largest in the lower left panel 
(however the uncertainties are also larger in that panel).

\section{Multi-parameter fits assuming a constant $\alpha_\mathrm{CO}$}

Tables~\ref{tab:estimators_all_aco3p2},~\ref{tab:estimators_no_ifu_aco3p2}, and~\ref{tab:estimators_no_12um_aco3p2} show the multi-parameter fit results
H$_2$ surface densities were computed assuming $\alpha_\mathrm{CO} =3.2 \> M_\odot \mathrm{(K\> km \> s^{-1} \> pc^2)^{-1}}$.

\begin{table*}
	\centering
	\caption{Same as Table~\ref{tab:estimators_all} but assuming $\alpha_\mathrm{CO} = 3.2 \> \mathrm{M_\odot} \mathrm{(K\> km \> s^{-1} \> pc^2)^{-1}}$.}
	\label{tab:estimators_all_aco3p2}
	\begin{tabular}{lllll|cc|ccc} %
		\hline
		& & & & & \multicolumn{2}{c|}{$\theta_i$ for pixel properties} & \multicolumn{3}{c|}{$\theta_i$ for global properties} \\
		\cmidrule(lr){6-7}  \cmidrule(lr){8-10} 
		RMS error & $n_\mathrm{gal}$ & $n_\mathrm{pix}$ & $\sigma_\mathrm{int}$ & Zero-point ($\theta_0$) &$\log \Sigma(\mathrm{12\>\mu m})$ & $(12+\log \mathrm{O/H})$ & $\log \Sigma_\mathrm{FUV}$ & $\log \Sigma_\mathrm{NUV}$ & $A(\mathrm{H\alpha})$  \\
		\hline
  $0.18$ &  $58$ &  $1126$ &  $0.16$ & $0.56 \pm 0.01$ & $0.73 \pm 0.01$ &  -- &  -- &  -- &  --  \\
  $0.16$ &  $30$ &  $573$ &  $0.14$ & $2.76 \pm 0.08$ & $0.85 \pm 0.01$ &  -- & $-0.21 \pm 0.01$ &  -- &  --  \\
  $0.14$ &  $27$ &  $552$ &  $0.14$ & $3.8 \pm 0.1$ & $0.95 \pm 0.01$ &  -- &  -- & $-0.30 \pm 0.01$ & $-0.15 \pm 0.01$  \\
  $0.14$ &  $27$ &  $552$ &  $0.14$ & $3.2 \pm 0.6$ & $0.94 \pm 0.01$ & $0.06 \pm 0.07$ &  -- & $-0.30 \pm 0.01$ & $-0.15 \pm 0.01$  \\
  		\hline
	\end{tabular}
\end{table*}

\begin{table*}
	\centering
	\caption{Same as Table~\ref{tab:estimators_no_ifu} but assuming $\alpha_\mathrm{CO} = 3.2 \> \mathrm{M_\odot} \mathrm{(K\> km \> s^{-1} \> pc^2)^{-1}}$.}
	\label{tab:estimators_no_ifu_aco3p2}
	\begin{tabular}{lllll|c|ccc} %
		\hline
		& & & & & $\theta_i$ for pixel properties & \multicolumn{3}{c|}{$\theta_i$ for global properties} \\
		\cmidrule(lr){6-6}  \cmidrule(lr){7-9} 
		RMS error & $n_\mathrm{gal}$ & $n_\mathrm{pix}$ & $\sigma_\mathrm{int}$ & Zero-point ($\theta_0$) &$\log \Sigma(\mathrm{12\>\mu m})$ & $\log \Sigma_\mathrm{FUV}$ & $\log \Sigma_\mathrm{NUV}$ & $A(\mathrm{H\alpha})$ \\
		\hline
  $0.18$ &  $58$ &  $1126$ &  $0.16$ & $0.56 \pm 0.01$ & $0.73 \pm 0.01$ &  -- &  -- &  --  \\
  $0.16$ &  $30$ &  $573$ &  $0.14$ & $2.76 \pm 0.07$ & $0.85 \pm 0.01$ & $-0.21 \pm 0.01$ &  -- &  --  \\
  $0.14$ &  $27$ &  $552$ &  $0.14$ & $3.8 \pm 0.1$ & $0.95 \pm 0.01$ &  -- & $-0.30 \pm 0.01$ & $-0.15 \pm 0.01$  \\
  $0.14$ &  $27$ &  $552$ &  $0.14$ & $3.5 \pm 0.1$ & $0.92 \pm 0.01$ & $-0.12 \pm 0.03$ & $-0.16 \pm 0.03$ & $-0.14 \pm 0.01$  \\
  		\hline
	\end{tabular}
\end{table*}

\begin{table*}
	\centering
	\caption{Same as Table~\ref{tab:estimators_no_12um} but assuming $\alpha_\mathrm{CO} = 3.2 \> \mathrm{M_\odot} \mathrm{(K\> km \> s^{-1} \> pc^2)^{-1}}$.}
	\label{tab:estimators_no_12um_aco3p2}
	\begin{tabular}{lllll|ccc|cc} %
		\hline
		& & & & & \multicolumn{3}{c|}{$\theta_i$ for pixel properties} & \multicolumn{2}{c|}{$\theta_i$ for global properties} \\
		\cmidrule(lr){6-8}  \cmidrule(lr){9-10} 
		RMS error & $n_\mathrm{gal}$ & $n_\mathrm{pix}$ & $\sigma_\mathrm{int}$ & Zero-point ($\theta_0$) &$\log \Sigma_*$ & $(12+\log \mathrm{O/H})$ & $\log \Sigma_\mathrm{SFR}$ & $\log \Sigma_\mathrm{NUV}$ & $b/a_\mathrm{disk}$ \\
		\hline
  $0.21$ &  $58$ &  $1126$ &  $0.20$ & $2.07 \pm 0.01$ &  -- &  -- & $0.50 \pm 0.01$ &  -- &  --  \\
  $0.20$ &  $42$ &  $942$ &  $0.18$ & $1.92 \pm 0.02$ &  -- &  -- & $0.49 \pm 0.01$ &  -- & $0.21 \pm 0.02$  \\
  $0.17$ &  $27$ &  $552$ &  $0.18$ & $0.8 \pm 0.1$ & $0.26 \pm 0.01$ &  -- & $0.30 \pm 0.01$ & $0.03 \pm 0.01$ &  --  \\
  $0.17$ &  $27$ &  $552$ &  $0.18$ & $-3.1 \pm 0.6$ & $0.23 \pm 0.01$ & $0.47 \pm 0.07$ & $0.32 \pm 0.01$ & $0.02 \pm 0.01$ &  --  \\
  		\hline
	\end{tabular}
\end{table*}

\bsp	%
\label{lastpage}
\end{document}